\documentclass[aps,pra,twocolumn,groupedaddress,showpacs]{revtex4}
\usepackage{graphicx,psfrag,bm,amsmath,amsfonts,amssymb,color,subfigure}

\newcommand{\be}{\begin{equation}}
\newcommand{\ee}{\end{equation}}
\newcommand{\beq}{\begin{eqnarray}}
\newcommand{\eeq}{\end{eqnarray}}
\newcommand{\ba}{\begin{array}}
\newcommand{\ea}{\end{array}}
\newcommand{\bea}{\begin{eqnarray}}
\newcommand{\eea}{\end{eqnarray}}
\newcommand{\ex}[1]{\mbox{e}^{#1}}

\newcommand{\eps}{\epsilon}
\newcommand{\im}[1]{\mbox{Im}\left[#1\right]}
\newcommand{\re}[1]{\mbox{Re}\left[#1\right]}

\newcommand{\besj}[2]{\mbox{J}_{#1}(#2)}
\newcommand{\beshp}[2]{\mbox{H}_{#1}^{+}(#2)}

\newcommand{\mcal}[1]{\mathcal{#1}}

\newcommand{\ra}{\rightarrow}

\newcommand{\g}{\gamma_{\perp}}
\newcommand{\gp}{\gamma_{\|}}
\newcommand{\om}{\omega}
\newcommand{\Om}{\Omega}

\newcommand{\lan}{\langle}
\newcommand{\ran}{\rangle}

\newcommand{\bma}{\begin{matrix}}
\newcommand{\ema}{\end{matrix}}

\newcommand{\Omu}{\Omega_{\mu}}
\newcommand{\psimu}{\Psi_\mu(\bm{x})}

\newcommand{\bx}{\bm{x}}
\newcommand{\bxp}{\bm{x}^{\prime}}
\newcommand{\tk}{\tilde{k}}
\newcommand{\yt}{\tilde{\kappa}_m}
\newcommand{\xt}{\tilde{q}_m}

\newcommand{\y}{y_m}
\newcommand{\x}{x_m}
\newcommand{\z}{z}
\newcommand{\delx}{\delta q_m}
\newcommand{\dely}{\delta \kappa_m}
\newcommand{\KM}{(n_0 k_m a)}
\newcommand{\KN}{(n_0 k_n a)}
\begin{document}
\title{Self-consistent multi-mode lasing theory for complex or random
lasing media}
\author{Hakan E. T\"ureci}
\email{tureci@phys.ethz.ch}
\affiliation{Institute of Quantum Electronics, ETH Zurich, 8093 Zurich, Switzerland}
\author{A. Douglas Stone and  B. Collier}
\affiliation{Department of Applied Physics, P. O. Box 208284, Yale University,
New Haven, CT 06520-8284, USA}

\date{\today}

\begin{abstract}
A semiclassical theory of single and multi-mode lasing is derived for
open complex or random media using
a self-consistent linear response formulation.  Unlike standard
approaches which use closed cavity solutions
to describe the lasing  modes, we introduce an appropriate discrete
basis of functions which
describe also the intensity and angular emission pattern outside the
cavity.  This constant flux(CF)  basis is dictated by the Green
function which arises when formulating the steady state Maxwell-Bloch
equations as a self-consistent linear response problem.  This basis
is similar to the quasi-bound state basis which is familiar in
resonator theory and it obeys biorthogonality
relations with a set of dual functions.  Within a single-pole
approximation for the Green
function the lasing modes are proportional to these CF states
and their intensities and lasing frequencies are determined by a set
of non-linear equations.  When a near threshold approximation is
made to these equations a generalized version of the
Haken-Sauermann equations for multi-mode lasing is obtained,
appropriate for open cavities.  Illustrative results from these
equations are given for single and few mode lasing states, for the
case of dielectric cavity lasers.  The standard near threshold
approximation is found to
be unreliable.  Applications to
wave-chaotic cavities and random lasers are discussed.
\end{abstract}
\maketitle

\section{\label{sectintro}INTRODUCTION}

A long-standing problem in laser theory is the formulation of a model
for lasing which
correctly treats the openness of the lasing medium/cavity and the
non-linearity of the coupled
matter-field equations.  This mathematical challenge has become of
great relevance with
the current high interest in complex or random lasers, for which the
mode geometry is not
fixed by the placement and orientation of mirrors and one typically
has multi-mode lasing behavior.   In this case many spatially complex
internal modes contribute to the external emission pattern and we
currently lack any theory to predict the directionality of emission
patterns and to understand and predict the output power as a function
of pump strength.  One class of systems of particular interest in
this regard are dielectric cavity lasers with complex and sometimes
chaotic ray dynamics \cite{Tureci05,SchwefelTureci04}; another class 
is the so-called
``random" lasers \cite{Cao03,Cao05} in which light undergoes diffusive
motion within the gain medium.  In the former case
the laser will typically have relatively high-Q modes whose lasing
properties are nonetheless determined by complicated competition
between spatially complex
modes.  The latter part of this work will focus on the dielectric cavity
case although the formalism we are developing should be
useful for both random lasers and for certain conventional lasers in which
mode competition is important.  We believe the current approach
solves the problem of treating the openness of the cavity in the
simplest possible manner, and makes clear the connection between the
resonances (or quasi-bound states) of the passive (cold) cavity and
the lasing modes of the active cavity.  In particular if a periodic
or multi-periodic solution of the Maxwell-Bloch (MB) equations
exists, then the lasing modes and frequencies are determined by a set
of self-consistent integral
equations the kernel of which is the
Green function of the linear problem with outgoing wave
boundary
conditions.  When the modes are relatively high-Q a simple
approximation to this
Green function implies that the spatial modes
are given by a certain set of states
which we refer to as constant flux (CF) states.  The effects of
spatial-hole burning are then described by interactions
between these modes, which unlike previous theories, are modes which exist
in all space and can thus be used to predict output power and
emission patterns from complex or random two and three-dimensional
laser cavities.   Throughout this work we employ the semiclassical
description of lasing implied by the MB equations, so the effects of
field quantization, such as quantum fluctuations, are not included.

The modal description of lasing assumes that the laser is in a regime
in which the steady-state
electric field in the pumped medium $\bm{E}(\bm{x},t)$ (which we will
henceforth refer to as the ``cavity") has a finite number of
frequencies and hence by Fourier transform can be seen as a sum of
non-linear modes $\bm{E}_{\mu}(\bm{x},\Omega_{\mu})$.  It is now well
known that lasers can have chaotic temporal dynamics which cannot be
described by a finite number of frequencies 
\cite{haken_light2,Lugiato92,ArecchiBR99,SunadaHI05}.
For clarity we wish to emphasize that we are not treating lasers which
exhibit dynamical chaos in this sense; the ``chaotic lasers" we are
treating are those with complex modal patterns which can be related
to the chaotic motion of light rays in the geometric optics limit.  A
term for this type of laser consistent with usage in the field of
quantum chaos is ``wave-chaotic". The temporal dynamics we are
treating is conventional multi-mode (MM) lasing, typical of the vast
majority of lasers.  There have been a number of interesting
experiments for which
the emission patterns from deformed (non-spherical or cylindrical)
dielectric cavity lasers have not been explainable in terms of
conventional ``whispering gallery" modes, but instead required
careful analysis of different modal patterns with different
ray-optical interpretations 
\cite{science98,GianordoliHSSFG00,gmachl02,rex02,sblee1,HarayamaFSI03,Chern03,PodolskiyNFC04}. 
Furthermore several experiments have found a dramatic variation of 
the output power for a
given pump strength with the shape of the laser cavity 
\cite{science98,GianordoliHSSFG00}. Our current formalism is proposed 
to describe
and predict the results of such experiments, which are inadequately
treated by standard approaches.

There are several methods to predict or describe lasing modes and
emission in this case of multi-periodic steady state solutions.  The
simplest approach is to work with the ``cold cavity"(CC); i.e. to find the
electromagnetic solutions for the Helmholtz equation describing light
in the cavity neglecting gain and non-linear effects.   If the lasing
modes have high Q values then the non-linear solutions are often
very similar spatially to these CC modes 
\cite{HarayamaDI03,HarayamaSI05} and from the experimentally observed 
MM lasing frequency spectrum one can associate the lasing state with 
a sum of cold cavity
modes (or a single mode if only one frequency is present).  Within this CC
approximation a further approximation is to treat the cavity as 
closed and work with hermitian
modes of a perfectly confining cavity. The Gaussian modes of
Fabry-Perot resonators shown in textbooks are solutions of this type.

Another more common approach within the CC approximation is to
approximate the lasing modes by the {\it quasi-bound(QB) states} or
resonances of the cavity.  These states are defined as the solutions
of the linear Maxwell wave equation (in the absence of gain) which
satisfy the boundary conditions at the cavity boundary and only have
outgoing waves at infinity. This outgoing wave boundary condition can
only be satisfied for discrete complex values of $k = q - i\gamma$
(with $\gamma >0$) which means that the outgoing spherical waves will
grow as $\exp[\gamma r] $ as $r \to \infty$.  Such modes are not
hermitian and are not orthogonal to one another (though a modified
orthogonality relation can be defined in {\it separable} non-hermitian
problems \cite{LeungLY94,ChingLvSTY98}). The quantity $\gamma$ 
determines the Q-value
of the resonance which can then be
used in formulating the laser theory. The Fox-Li method \cite{foxli1960} dating
from the early days of laser theory is one
technique for finding such quasi-bound modes; there are a number of
modern methods for doing this as well 
\cite{manenkov94,noeckel_thesis,HentschelR02,Wiersig03a,Tureci05}. 
This theoretical approach has been widely applied to analyze post 
facto the modes of wave-chaotic
or random lasers (see \cite{Tureci05} for references). Its obvious 
drawback is that the
method is not predictive.  First, a larger set of QB modes is found
and a subset, the lasing modes, are chosen as those which "look" like
the experimental results.  Furthermore, there is no means within this
approach to understand and predict the output power of the laser.

The standard approach to go beyond the CC approximation is to use the
simplest semiclassical lasing theory, describing a pumped medium of
two-level atoms coupled to light within a cavity, known as the
Maxwell-Bloch (MB) equations.  Our theory below is based on analysis
of these MB equations. The MB equations can be analyzed to find
single-frequency solutions,
usually approximated by single modes of the cold cavity.  One then
finds that it is the highest-Q cold cavity mode which lases first and
that its frequency is shifted from the cold cavity frequency towards
the atomic transition frequency but not by a large amount if the Q of
the cavity is sufficiently high 
\cite{haken_light2,lamb_book,HarayamaFSI03,HarayamaDI03}.  The 
spatial field distribution
is typically assumed unchanged from the CC mode which can be
approximated by the closed cavity mode or the corresponding CC
resonance.  The complication in the theory comes when one attempts to
describe multi-mode lasing.  First the general non-linear MB
equations cannot be solved analytically  using a modal expansion of
the electric field.  Exact numerical solution of the equations can be
done for some cases and is useful \cite{HarayamaSI05}; however this
becomes computationally intractable in the short wavelength limit of
interest here and it is difficult to extract qualitative physical
ideas from such an approach.   A nice method due to Haken \cite{haken63}
dates back to the early days of laser theory; one limit of
our theory is an extension of this approach. The Haken method
expands the electric field which solves the MB equations as a sum of
CC modes and writes an equation of motion for the amplitude of each
mode which can be reduced to a (constrained) linear  equation for the
modal intensities in the near threshold approximation.
The major drawback of the Haken approach is that it is formulated in
terms of the modes of the ideal closed cavity and calculates only the
internal field intensities of these modes; thus without some ad hoc
assumptions it doesn't predict output power or directional emission
patterns.  This is done to exploit the orthogonality of closed cavity
modes which simplifies the mathematical description. For comparison
to our results below, the Haken near threshold MM lasing equations
for the steady state electric field intensities $I_{\nu}$ are
\be
1-\frac{\kappa_\mu}{D_0} = \sum_{\nu}g(\Omega_\nu) A_{\mu\nu} I_\nu
\label{eqdiagnin1}
\ee
where $\kappa_{\mu}, D_0, g(\Omega_\nu)$ are the cavity decay rate,
pump strength, and laser
gain profile respectively (in appropriately
scaled units) and
\be
A_ {\mu\nu}=\int d^3 \bx\,|\varphi_\mu(\bx)|^2 |\varphi_\nu (\bx)|^2
\ee
is a matrix which describes interactions between CC modes
$\varphi_{\mu}(x), \varphi_{\nu}(x)$.
The complexity of modal
solutions in space thus determines the lasing
amplitudes through the properties of
this matrix.  This equation was derived by Haken and Sauermann in
1963 to describe the effects of spatial hole-burning: the fact that
lasing modes deplete the inversion in a spatially-varying manner
which can then allow many  modes to lase in steady-state when the
pump is sufficiently strong. A statistical analysis of certain
aspects of these equations for the case of closed cavities
described by random matrix theory was performed by Misirpashaev and
Beenakker some time ago \cite{misirpashaevB98}.

Note that while Eq.~\ref{eqdiagnin1} appears to be a simple
inhomogeneous linear equation for the $\{I_\nu\}$, it cannot be
trivially inverted to yield these intensities since we have the
additional constraint that $I_\nu \geq 0, \forall_\nu$.  The equation
thus needs to be solved by an iterative procedure.  In a recent work
\cite{tureci05b} we describe this procedure and used it along with some {\it ad
hoc} assumptions to calculate multi-mode output intensities for
deformed dielectric cylinder lasers, making contact with the
experimental results of Ref.~\cite{science98}.

In the current work we present a formalism which gives a generalization of
Eq.~\ref{eqdiagnin1}, and which avoids these {\it ad hoc} assumptions.
The basic innovation of the approach is to formulate the solution of
the MB equations as a self-consistent linear response problem and
write this solution in terms of the Green function of the wave
equation with the boundary condition of constant outgoing flux at
infinity.  Using this Green function the lasing solutions and
frequencies are
determined by an integral equation which can be used
to describe multimode lasing
in both high-Q and low-Q cavities, and
in random media (acting as a cavity).
To describe complex cavities with high-Q modes it is useful to
represent this Green function in terms of a new set of
linear eigenfunctions which satisfy the wave equation with real
wavevector
but purely outgoing boundary conditions; we refer to these
functions as constant flux (CF)
states. They are similar to the resonant solutions of the cold cavity
in that they have a complex wavevector inside the lasing
medium, but they differ in that they have a real wavevector outside
the medium and hence give a well-defined field at infinity.  Like the
quasi-bound states, the  CF states are not orthogonal, and the Green
function
involves both the CF states and their biorthogonal partners
which represent states of constant
incoming flux from infinity.  The use of biorthogonal pairs of
resonator states is well established in resonator theory, and it well is
known that the two states correspond to different directions of
propagation through the apparatus (see Ref.~\cite{siegman_book}, pp 
847-857).  Our
CF states are just a generalization of this idea.  The new feature
here is that a simple ``single-pole"
approximation to the Green
function implies that the lasing modes are proportional to
the CF
states and leads to a generalization  of the
Haken-Sauermann
multi-mode lasing equations in terms of
CF states.  Iterative
solutions of these equations predict the multi-mode lasing
states
both inside and outside the cavity and hence provide a
predictive theory of output
power and directional emission from
complex cavities.

The paper is organized as follows.  In Section~\ref{sectselfconsle} 
we derive the
self-consistent integral equation for the lasing modes assuming that
a multi-periodic solutions exists.  To do this we introduce a Green
function for the inhomogeneous wave equation
within the cavity.  In
Section~\ref{sectbiocfstates} we discuss the boundary conditions on this Green
function and its spectral representation.
  We show that in order to
satisfy the outgoing wave boundary conditions this spectral
representation is expressed in
  terms of two sets of biorthogonal
functions (the CF states).  The approximation
corresponding
  to standard multi-mode lasing theory is to approximate
the Green function by the contribution of the single-pole nearest
  to the
lasing frequency.  This implies that the lasing modes are proportional to
a single CF state.  We discuss examples of CF
  states for a slab
(1D), cylinder (2D) and deformed cylindrical dielectric cavity.  In
Section~\ref{sectgenmmeqns} we derive the multi-mode
  lasing equations which follow
from this approximation, leading to linear and non-linear
generalizations of the
  Haken-Sauermann equations, valid for the {\it
open} cavity.  In Section~\ref{sectitsolgenhseq} we discuss iterative 
methods for solving
these equations for the modal intensities and the lasing frequencies
and then present solutions for the single and two mode cases, beyond
the standard Haken-Sauermann near threshold approximations.  Our
results suggest that the near threshold
  approximation introduces
significant error and greatly overestimates the number of lasing
modes at a given pump
  power.  We summarize our results in Section~\ref{sectconcl}
and provide further
  detail on CF states in the Appendices.

\section{Derivation of the self-consistent semiclassical laser equations}
\label{sectselfconsle}
Following the standard semi-classical laser theory
\cite{haken_light2,haken63}, we start with the Maxwell-Bloch
equations (MB) in the form
\be
\nabla^2 E - \frac{1}{c^2}\ddot{E} = \frac{4\pi}{c^2} \left[
\ddot{P^+} + (\ddot{P^+})^*\right]
\label{eqinwave}
\ee
\bea
\dot{P^+} &=& -(i\omega_a + \g ) P^+ + \frac{g^2}{i\hbar} ED \label{eqpol1}\\
\dot{D} &=& \gp (D_0 - D) - \frac{2}{i\hbar} E
\left[(P^+)^*-P^+\right] \label{eqinv1}
\eea
This is a set of non-linearly coupled spatio-temporal partial
differential equations for the electric field amplitude
$E(\bm{x},t)$, the macroscopic polarization $P(\bm{x},t)=n_a
(g\rho_{12}(\bm{x},t) + g^* \rho_{21}(\bm{x},t)) = P^+ + P^-$ and the
inversion $D(\bm{x},t)=n_a(\rho_{22}(\bm{x},t)-\rho_{11}(\bm{x},t))$.
Here, $P^+ = (P^{-})^*=n_a g \rho_{12}$.   The parameters entering 
the equations are as follows: $D_0$ is the
external pump strength, 
$g$ is the dipole moment matrix element, $\gamma_{\perp}$ and $\gp$ are
phenomenological damping constants for the polarization and the
inversion, respectively, and $c$ is the speed of light.   Note that
$E$, $D$ and $P$ are real valued fields.

We will make the following
assumptions:
\begin{itemize}
\item We focus here on a scalar field $E$ defined in two
space-dimensions $\bm{x}=(x,y)$. For instance, in the case of a
dielectric cylinder laser with arbitrary cross-section, $E$ will
denote the $z$-component of the electric or magnetic field for $k_z =
0$ modes (see Ref.~\cite{Tureci05} for the justification of this model in
the case of a dielectric cylinder laser).
\item We assume a uniform, homogeneously broadened atomic medium with
density of atoms $n_a$, atomic transition frequency $\omega_a$ and
the quantum mechanical density matrix $\rho_{ij}(\bm{x},t)$.
\end{itemize}

We decompose $P = P_L + P_{NL}$, into a linear component $P_L$ (for
instance the non-resonant response of the substrate material), and a
non-linear resonant component $P_{NL}$ due to the gain medium, here
taken to be the uniformly distributed set of two-level atoms. Let
\begin{equation}
P_L(\bm{x},t) = \int dt' \chi(\bm{x},t-t') E(\bm{x},t')
\end{equation}
Decomposing the fields $P=P_L+P_{NL}$ and $E$ as
\begin{equation}
F(\bm{x},t)=F^+(\bm{x},t)\ex{-i\omega_a t } + F^-(\bm{x},t)\ex{i\omega_a t }
\label{eqsve1}
\end{equation}
where $F=(E,P_L,P_{NL})$ and using the {\em slowly varying envelope
approximation} and the {\em rotating wave approximation},
Eq.~\ref{eqinwave} becomes
\begin{equation}
\nabla^2 e + \frac{\om_a^2}{c^2} (e + 4\pi p_L) +
\frac{2i\om_a}{c^2}(\dot{e} + 4\pi\dot{p}_L) = -
\frac{4\pi\om_a^2}{c^2} p_{NL} \label{eqMB000}
\end{equation}
The envelope equations corresponding to 
Eqns.~\ref{eqinwave}-\ref{eqinv1} become
\bea
\partial_t ( \hat{n}^2 e ) &=& \frac{i}{2\om_a} \left[ \om_a^2
\hat{n}^2 + c^2 \nabla^2 \right] e + 2i\pi\om_a p_{NL} \label{eqMB0}\\
\dot{p}_{NL} &=& -\g p_{NL} + \frac{g^2}{i\hbar} eD  \label{eqMB1}\\
\dot{D} &=& \gp\left( D_0-D \right) - \frac{2}{i\hbar}\left( ep_{NL}^* -
p_{NL}e^* \right) \label{eqMB2 }
\eea
Here, $p_{NL}(\bm{x},t) = P_{NL}^+(\bm{x},t)$, $p_L(\bm{x},t) =
P_{L}^+(\bm{x},t)$, $e(\bm{x},t) = E^+(\bm{x},t)$, and $\hat{n}^2$
stands for the convolution  operator $\int  dt' \, n^2
(\bm{x},t-t')$. The time-dependent index of refraction $n(\bm{x},t)$
is related to $\chi(\bm{x},t)$ by $n^2(\bm{x},t)=1+4\pi\chi(\bm{x},t)$.
In the derivation of these equations we have assumed that the
residual time-dependence of the envelopes is much slower than $\om_a$
and that $\om_a \gg \g$ and dropped the terms proportional to
$\ddot{e}, \ddot{p}_{NL}, \om_a \dot{p}_{NL}, \ddot{p}_L$.

The properties of the index of refraction $n(\bm{x},t)$ of the
background medium
will depend on the particular complex lasing
system under investigation; for dielectric
cavity lasers it will be
constant in the cavity and unity outside, for a random laser
it will vary randomly in space within some region and then fall off to unity
at the edges of the medium.  For the wave-chaotic
dielectric cavity the ``randomness" comes from the scattering at an
irregularly shaped boundary. We will also replace $p_{NL}$ by $p$
henceforth.

We assume now that the fields
$p(\bm{x},t)$ and $e(\bm{x},t)$ are multi-periodic in time
\begin{equation}
p(\bm{x},t) = \sum_\mu p_\mu(\bm{x}) \ex{-i\Omega_\mu t} \quad , 
 \quad e(\bm{x},t) = \sum_\mu \Psi_\mu(\bm{x})
\ex{-i\Omega_\mu t} \label{eqmmansatz}
\end{equation}
in the {\em steady state}. In contrast to previous approaches 
\cite{haken_light2,lamb_book,LugiatoOTNP90,StaliunasTW93,MandelEO93,Zehnle98}, 
we leave the spatial functions $p_\mu(x)$ and
$\Psi_\mu(x)$ to be {\em unknown} functions and $\Omega_\mu$ to be
the unknown lasing frequencies to be determined. Such a solution with
a finite number of discrete frequencies (a ``multi-mode'' lasing
solution) is only possible if the inversion is approximately
stationary \cite{FuH91}. Henceforth, we will assume this to be the
case. From Eq.~(\ref{eqMB1}), the ansatz Eq.~(\ref{eqmmansatz}) and the
time-independence of the inversion we obtain
\begin{equation}
p_\mu(\bx) = \frac{g^2}{i\hbar} \frac{D(x)}{-i\Omega_\mu+\g} \Psi_\mu (\bx)
\end{equation}
and
\begin{equation}
D(\bx) \approx \frac{D_0}{1+\frac{1}{e_c^2}\sum_{\mu} g(\Omega_\mu)
|\Psi_\mu(\bx)|^2 }
\end{equation}
Here, $g(\Om) = \g^2/(\Om^2 + \g^2)$ is the laser linewidth and $e_c
= \hbar\sqrt{\g\gp}/2g$ gives the typical electric field scale. 
Henceforth we will measure
the lasing modes and polarization, 
$\Psi_\mu  (\bx), p_\mu(\bx)$ in units of $e_c$.  We
can then rewrite Eq.~\ref{eqMB0} in the form
\begin{equation}
\left[ 2i\om_a \partial_t \, \hat{n}^2 + \om_a^2 \hat{n}^2 + c^2
\nabla^2 \right] e(\bm{x},t) = -4\pi\om_a^2 p(\bm{x},t)
\label{eqWEwithsource} \\
\end{equation}
Substituting the results for $p_\mu (\bx)$ into Eq.~(\ref{eqWEwithsource}),  the pumped atomic medium  becomes equivalent
to a {\em multi-periodic forcing term} on the right side of the
Eq.~(\ref{eqWEwithsource}), given by
\begin{align}
-4\pi\om_a^2 p(\bm{x},t) = -4\pi\om_a^2 & \frac{g^2}{i\hbar}
\frac{D_0}{1+  \sum_{\nu} g(\Omega_\nu) |\Psi_\mu(\bx)|^2
} \nonumber \\
& \times \sum_\mu \frac{1}{-i\Omega_\mu + \g} \Psi_\mu(\bx) \ex{-i\Omega_\mu t}
\end{align}
where the polarization which is the source for $e(\bm{x},t) =
\sum_{\mu} \Psi_\mu (\bx)
e^{-i\Om_\mu t}$ is itself a non-linear function of $ \{\Psi_{\mu}
\}$.
Thus Eq.~(\ref{eqWEwithsource}) gives a self-consistent set of
equations for the mode amplitudes $\Psi_\mu(\bm{x})$ and lasing
frequencies $\Om_\mu$.

To derive from Eq.~(\ref{eqWEwithsource}) an
integral equation for each $\Psi_\mu(\bm{x})$,
we first Laplace transform Eq.~(\ref{eqWEwithsource}) to the frequency domain:
\begin{equation}
(\om_a + \om)^2n^2(\om)\tilde{e}(\bm{x},\om) + c^2\nabla^2
\tilde{e}(\bm{x},\om) = -4 \pi \omega_a^2 \tilde{p} (\bm{x},\om)
\label{eqwaveeqnwsrc}
\end{equation}
Here, $n^2(\om) = 1 + 4\pi \tilde\chi(\om)$, the Laplace transforms
are defined by $\tilde{p}(\om) = \frac{1}{2\pi}\int_{0}^{\infty} dt
\ex{i(\om + i \epsilon) t} p(t)$, and
we only retain the
infinitesimal imaginary part of the frequency, $ \epsilon$, when it
is needed for convergence. Note that the real frequency variable
$\om$
corresponds to the {\it residual} time-dependence of $p(x,t)$
after removing the rapid variation at frequency $\om_a \gg \omega$,
and hence we have used $\om_a^2 + 2\om\om_a \approx (\om_a+\om)^2$.

Initially we treat the right hand side of Eq.~(\ref{eqwaveeqnwsrc}) 
as a given source
and introduce a Green function for the inhomogeneous equation, which
ensures that the solution $\tilde{e}(x,\om)$ satisfies the
appropriate boundary conditions, i.e. that the field has
only outgoing contributions.  Assuming for the moment that such a
Green function exists we can immediately write a formal solution to
Eq. (\ref{eqWEwithsource}) {\it within the cavity}
\begin{align}
\tilde{e}(\bx,\om) & = \sum_{\mu} \frac{\Psi_\mu (\bx)}{2\pi i(\om -
\Om_\mu + i \epsilon)} \nonumber \\
& =-\frac{4\pi\om_a^2}{c^2} \int_{cavity} d\bxp  \,
G(\bx,\bxp |\om) \tilde{p}(\bxp,\om)
\label{eqGreenFcnOm}
\end{align}
The integral on the rhs of Eq. (\ref{eqGreenFcnOm})
is only over the cavity because
the source (pumped medium) is zero
outside of the cavity.  The cavity Green function
satisfies the equation
\begin{equation}
[\nabla^2  + \frac{n^2 (\om)}{c^2} (\om + \om_a)^2]
G(\bm{x},\bm{x}'|\om) = \delta^3(\bm{x}-\bm{x}')
\end{equation}
which formally inverts Eq.~(\ref{eqwaveeqnwsrc})  to yield Eq.~(\ref{eqGreenFcnOm}) for
the electric field component
$\tilde{e}(\bx,\om)$ for $\bx$ within
the cavity.   Outside the cavity the wave equation changes
and a
different Green function would be needed to invert the equation.
However as the cavity Green
function determines the solution on the
boundary of the cavity, one can just use the outgoing wave
boundary
condition to fix the solution outside.  This will be done through our
definition of
the CF states below.

Both sides of Eq.~(\ref{eqGreenFcnOm}) involve sums over the modes $\mu$
weighted by the factor $1/2\pi i(\omega -\Om_\mu + i\epsilon)$ which
will give the multi-periodic time dependence of the solution when it
is inverse Laplace transformed.  The analytic structure
of the relevant  Green function can be shown to produce no additional
harmonic dependence in time as $t \to \infty$ so that it is possible
to equate the residues of these real poles to yield an integral
equation for each of the modes $\psimu$:
\begin{align}
\psimu =
 & i \frac{4\pi \omega_a^2 g^2 D_0}{\hbar c^2  (-i \Omu +\gamma_{\perp})} \nonumber \\
& \times \int_{cavity}
d\bxp \frac{ G(\bx,\bxp | \Om_\mu) \Psi_\mu (\bxp)}{1 +
\sum_\nu g(\Om_\nu)
      |\Psi_\nu (\bxp) |^2}
\label{eqscint}
\end{align}
This set of non-linear integral equations for the lasing modes is
completely general assuming a multi-periodic solutions exists.
If the modes of the cavity are sufficiently high-Q that the imaginary 
part of their
frequency is much smaller than the typical mode spacing
then one can introduce an approximation for the Green function
of Eq. (\ref{eqscint}) which leads to generalizations of the
conventional multi-mode lasing equations (see below).
However, it is possible to work directly with Eq. (\ref{eqscint})
when, as in a
random "diffusive" cavity, the assumption of
well-separated high-Q modes
is not correct. In this case different
approximations to the Green function,
similar to those made for disordered electronic systems, would be
more appropriate and would yield composite lasing
modes generated from many very broad cold cavity resonances.  We will
not pursue this direction in the current work, but wish to point out
the feasibility of such an approach.

\section{Biorthogonal Constant Flux States}
\label{sectbiocfstates}
\subsection{General Definition and Properties}

To express the lasing solutions $\psimu$ as an expansion in a set
of linear cavity modes we now introduce a spectral representation of
the Green function $G(\bx,\bxp | \omega)$ of the form:

\begin{equation}
G(\bx,\bxp |\om) = \sum_m
\frac{\varphi_{m}(\bx,\om)\bar{\varphi}^*_{m}(\bxp,\om)}{ n^2 [(k +
k_a)^2- (k_m + k_a)^2]}
\label{eqspecrep}
\end{equation}
where $k=\om /c,k_a = \om_a/c$ and  the functions in the numerator
are chosen so that G satisfies the
outgoing wave boundary conditions.  It is here that the non-hermitian
nature of these boundary conditions requires a change from the
standard approach utilizing closed cavity (hermitian) modes.  We
needed to introduce in Eq.~(\ref{eqspecrep})  {\it two} sets of 
functions which are
{\it biorthogonal} \cite{morse&feshbach} $\{\varphi_{m}(\bx,\om)\},
\{\bar{\varphi}_{m}(\bx,\om)\}$.   The set $\{\varphi_{m}(\bx,\om)\}$
satisfies the eigenvalue equation
\begin{equation}
-\nabla^2 \varphi_m(\bm{x},\om) = n^2 (k_m + k_a)^2
\varphi_m(\bm{x},\om)
\label{eqcfgenev}
\end{equation}
defined in the cavity, $\mcal{D}$.  While the Green function of 
Eq.~(\ref{eqspecrep})
is only used to find the electric field inside the cavity (and
would give an incorrect answer for the electric field outside), one
can define the functions $\varphi_m$ outside the cavity as well.
Outside the cavity they satisfy the free wave  equation
with the
fixed external wavevector, $k_a + k$, (not $k_a +
k_m$):
\begin{equation}
-\nabla^2 \varphi_m(\bm{x},\om) = (k_a + k)^2
\varphi_m(\bm{x},\om),
\end{equation}
with the outgoing wave boundary conditions
expressed in d-dimensions by:
\begin{equation}
r^{(d-1)/2} \varphi_m (r\rightarrow\infty,\om) \sim e^{i ( k_a + k)r},
\label{eqoutgoingCF}
\end{equation}
equivalently it can be expressed as a linear homogeneous boundary
condition on the function $\Phi_m (\bx) \equiv r^{(d-1)/2} \varphi_m
(r)$ in the form $d\ln  \Phi/dr =  i (k_a + k) $ as $ r \to \infty$.

Thus each $\varphi_m$  satisfies a {\it different} differential
equation inside and outside the cavity but are connected by
the continuity conditions
\begin{equation}
\varphi_m |_{\partial \mcal{D}^-} = \varphi_m |_{\partial \mcal{D}^+}
\quad \text{,}\quad \partial_{\bm{n}}\varphi_m |_{\partial
\mcal{D}^-} = \partial_{\bm{n}} \varphi_m |_{\partial \mcal{D}^+}
\label{eqcontBC0}
\end{equation}
at the boundary of the cavity (here $\partial\mcal{D}$ denotes the
boundary of $\mcal{D}$ and $\partial_{\bm{n}}$ is the normal
derivative on $\partial\mcal{D}$).  This choice guarantees that in
the approximation developed below, in which
each lasing mode is
proportional to a single $\varphi_m$, the electric field will also be
continuous at the dielectric
boundary, satisfy the free wave
equation outside, and have only outgoing components at infinity.  The
spectral
representation, Eq.~(\ref{eqspecrep})  is only meaningful for $\bx,\bxp$
within the cavity and on its boundary.
We will focus on the typical
case in which one can impose the outgoing wave boundary condition
directly on the cavity boundary (as in the slab and cylinder examples
given below). In
this case the eigenvalues $(k_m + k_a)^2$  are determined by linear
homogeneous boundary
conditions on a finite region, and are complex
due to the complexity of the logarithmic derivative at the boundary.

Note however, that this condition differs subtly
from the condition defining the quasi-bound (QB) states, for which 
the complex eigenvalue $k_m$ 
itself would appear in the logarithmic 
derivative at the boundary, and hence in
the external outgoing wave.  As a result, as noted earlier, the QB
states grow exponentially at infinity and carry infinite flux
outwards.  The states $\varphi_m$ we have just defined have real
wavevector at infinity and carry constant flux; hence we refer to
them as constant flux (CF) states.

The CF states have appeared naturally from the condition
that the Green function of Eq.~(\ref{eqGreenFcnOm}) satisfy the 
correct radiation
boundary condition; this allows
us to formulate a lasing theory valid for an open cavity in terms of
these states. In the
simplest approximation, to be developed below, the lasing modes {\it 
are} just single CF states
(those above threshold for a given pump power), each scaled by an overall
intensity factor determined by solving Eq.~(\ref{eqscint}), or
its near-threshold approximation, which turns out to be
a generalization of the Haken-Sauermann equations.
Note that CF states have complex wavevector inside the cavity, and 
are amplified
while traveling in the gain medium,
but real wavevector and conserved
flux outside as one expects for lasing modes (see Fig 
\ref{figdislabQBCF} below).  We show in the
Appendix that for high-Q modes the CF wavevector inside the medium
is very close to that of the corresponding QB state, justifying the
use of QB states to calculate the Q value of lasing modes, as is
often done.

As mentioned already, due to the non-hermitian
nature of the associated eigenvalue problem we need to introduce a second
set of functions, which satisfy different boundary
conditions, in order to represent the Green function of 
Eq.~(\ref{eqspecrep}).  The
functions $ \{ \bar{\varphi}_m(\bm{x},\om)\}$ satisfy
\begin{equation}
-\nabla^2 \bar{\varphi}_m(\bm{x},\om) =
n^*(\om)^2 (\bar{k}_m + k_a)^2 \bar{\varphi}_m(\bm{x},\om)
\label{eqcfadjointgen}
\end{equation}
with the continuity conditions (\ref{eqcontBC0}) but with the {\em
incoming} wave boundary condition
\begin{equation}
r^{(d-1)/2} \bar{\varphi}_m (r\rightarrow\infty,\om) \sim e^{-i (k_a + k)
r}. \label{eqadjbcsatinf}
\end{equation}
It can easily be shown \cite{morse&feshbach} that
$\bar{k}_m^2= k_m^{2*}$
but that $\bar{\varphi}_m(\bm{x},\om) \neq
\varphi_m^*$ in general.     (We will see in an example below that when
$\bar{\varphi}_m(\bm{x},\om)$ is real for real frequencies,
then
$\bar{\varphi}_m(\bm{x},\om)$ does equal
$\varphi_m^*$).

The key property which motivates the use of the two
sets of functions
in the spectral representation is that they
satisfy biorthogonality
with respect to the following inner product
\begin{equation}
\lan\lan \varphi_m | \varphi_n \ran \ran =  \int_\mcal{D} d^2\bm{x}
\, \bar{\varphi}^*_m(\bm{x},\om)\varphi_n(\bm{x},\om) = \eta_m
(\omega) \delta_{mn}.
\label{eqbiorthdp1}
\end{equation}

The normalization factor $\eta_m$ can be set equal to unity for any
specific choice of the frequency $\omega$, but it is useful to retain
it explicitly in some contexts to denote that the normalization
depends on $\omega$.  Note that we have here the usual hermitian inner
product, except that it is defined between eigenfunctions living in
adjoint spaces. These
pairs of functions with eigenvalues related by complex conjugation
are called biorthogonal pairs or partners.

The boundary condition (\ref{eqadjbcsatinf})  clearly has the meaning
that the adjoint CF states have constant {\it incoming flux} at infinity.
For these states one can view the source as being at infinity and
emitting radiation which impinges on the dielectric medium and then
decays inwards at the same rate as the CF states grow while moving
outwards from the origin.  In other words the dielectric medium acts
as an amplifying medium for outgoing CF states and as an absorbing
medium for ingoing (adjoint) CF states.

\subsection{CF states and
multi-mode lasing}

The importance of CF states becomes clear when we
consider the self-consistent integral
equations for the lasing modes
(\ref{eqscint}) using the CF spectral representation of the
Green
function.   We can write Eq. (\ref{eqspecrep})
in the form:
\begin{equation}
G(\bx,\bxp |\om) = \frac{c^2}{n^2} \sum_m
\frac{\varphi_{m}(\bx,\om)\bar{\varphi}^*_{m}(\bxp,\om)}{ 2 \om_a
(\om -\om_m)}
\label{eqspecrep2}
\end{equation}
where we have used the fact that $\om_a \gg
\om,\om_m$.  As already noted, the $  \{k_m = \om_m (\om)/c \equiv q_m - i
\kappa_m \}$ are complex but for the high-Q modes of interest their
imaginary parts, $\kappa_m$, will be
small compared to the typical
spacing between their real parts.
Therefore the Green function is
going to be multiply-peaked as a function of $\om$ (which
is real) around the values $\om \approx c q_m$.   The $m^{th}$ term
in the spectral sum is large
when $\om = c q_m \equiv w_m$,
proportional to $1/\kappa_m$; whereas the other terms will have
denominators
at least as large as $\delta q$, the mean spacing
between the real part of the eigenfrequencies (which is
roughly the
same as the mode spacing of the closed cavity).  This spacing is
constant in 1D, and
only decreases as some
power of the parameter
$kR$ ($R$ is the typical linear size of the resonator) in higher D,
whereas the imaginary
quantities $\kappa_m$ can be exponentially
small in this parameter.  For typical dielectric cavity
resonators
there are many high-Q modes for which $\kappa_m \ll \delta q$ and the
Green function
will be dominated by the single nearest CF pole at
$\om = w_m$.  It is thus natural to
introduce the single-pole
approximation to the CF Green function in which we replace the
full
Green function near the lasing frequencies by the single term
involving the nearest pole in the spectral
representation
\footnote{We note that this approximation breaks down
for
degenerate or quasi-degenerate poles. In such a situation, one has to
take into account the contribution from more than one pole and a
consistent multi-pole theory can be derived leading to correct
treatment of collective frequency locking phenomena}.

Thus we
assume that the possible lasing frequencies are in one-to-one
correspondence with
the real parts of the CF eigenvalues $\{w_m = c
q_m\}$.
We write $\om_m
\approx \Om_\mu + \delta \om_m^{\mu}
- i  \kappa_m^{\mu}$ ($\kappa_m^{\mu}>0$).
In the single-pole
approximation, the lasing modes are given by
$\Psi_\mu (\bx) = a_m^{\mu} \varphi_m^\mu (\bx)$ and are just
proportional to single CF states.
Note that not all CF states will
lase; the solution Eq. (\ref{eqscint}) will find that many of
the coefficients $a_m^{\mu}$ are zero; the remainder will give the
intensities of the various lasing modes.
Thus, as noted above, the
lasing modes at this level of approximation are
identical to (a subset of) the CF states up to an overall scale
factor.  We will work within this approximation for the remainder of
this article, so we can now adopt a simpler notation, dropping the
subscript $m$: $a_m^\mu \to
a_\mu, \varphi_m^\mu (\bx) \to \varphi_\mu (\bx)$.

Having just shown that CF states are the correct functions to
describe
solutions for high-Q lasing modes of open resonators, before
we discuss the derivation and solution of the lasing equations for
their intensities and frequencies,
we will examine some examples of
CF states and relate them to the more familiar QB states.

\subsection{Examples of CF states}
\subsubsection{\label{sect1dCF}One-Dimensional CF
state}

\begin{figure}[hbt]
\includegraphics[clip,width=0.45\linewidth]{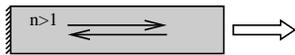}
\caption{Schematic of a dielectric slab cavity of index $n_0$ and 
length $a$ with perfectly reflecting mirror at the origin.}
\label{figdislabschem}
\end{figure}

The simplest example of a set of CF states which also has the
virtue of being essentially exactly solvable are the states of a
semi-infinite slab laser (see Fig.~\ref{figdislabschem}).  Note that 
this example is also a crude model for an
edge-emitting semiconductor laser
cavity.
Consider a dielectric medium
of index $n=n_0$ which is uniform and infinite in the $y,z$
directions and extends from $x=0$ to $x=a$ in the $x$ direction.  At
$x=0$ it is terminated in a perfectly reflecting mirror and from
$x=a$ to $\infty$
there is vacuum $(n=1)$. To clarify the difference
between the CF states and the usual cold cavity resonances, first
let us consider the quasi-bound states of such a resonator.  These
are defined as the solutions of
\begin{equation}
-\partial^2_x \varphi_m (x) = n_0^2 \tilde{k}_m^2 \varphi_m(x)
\end{equation}
and the boundary condition (which can moved to the interface)
$\partial_x \varphi_m(x =a) = i \tk_m \varphi_m(x =a) $.
The complex eigenfrequencies of this problem are the solution of
\begin{equation}
\tan (n_0 \tk_m a) = - in_0
\end{equation}
which can be explicitly solved to yield
\begin{equation}
n_0 \tk_m a = \pi (m + 1/2) -i
\frac{1}{2}\ln[\frac{n_0+1}{n_0 - 1}] \equiv n_0 (q_m -
i\kappa_0)a ,
\label{eq1dslabquant1}
\end{equation}
where $m=1,2, \ldots, \infty$.  Note that the
imaginary part of the wavevector is always negative,
as it should be,
and in this case is constant, corresponding to the fixed
transmissivity of a dielectric interface of index $n_0$ at normal
incidence.  Since these resonances are outgoing at infinity, each QB
state varies as $e^{+i \tk_m x} \propto e^{+\kappa_0 x}$ and grows
exponentially at infinity as mentioned in the general discussion of
QB states in the introduction.

Now consider the CF states; inside the medium they satisfy
exactly the same equation and the continuity conditions at the
dielectric interface are the same as well, but the CF state {\it
outside} the dielectric satisfies the wave equation with a fixed
external wavevector:
\begin{equation}
-\partial^2_x\varphi_m (x) =  k^2 \varphi_m(x),
\end{equation}
and with the corresponding outgoing wave boundary
condition, $\partial_x \varphi_m(x =a ) = i k \varphi_m(x =a)$.
This defines a {\em family} of basis states  depending on the
choice of the real external frequency $\om = ck$.
These eigenvalues satisfy the equation:
\begin{equation}
\tan (n_0 k_m a) = - in_0
\frac{k_m}{k}.
\end{equation}
In the Appendix it is shown that this
is a well-defined eigenvalue problem for any value of
$k$  and the eigenvalues are always complex with negative imaginary
part, similar
to the usual QB states.  But again, unlike the QB states, since $k$
is real, these states simply oscillate
at infinity as $e^{ikx}$ and carry a constant outgoing flux.

\begin{figure}[hbt]
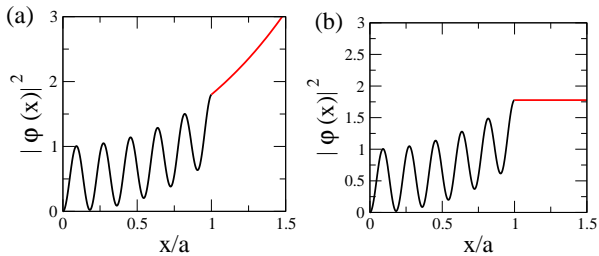

\includegraphics[clip,width=0.45\linewidth]{wfplotQBslab.eps}
\includegraphics[clip,width=0.45\linewidth]{wfplotslab.eps}
\caption{Solutions of the QB and CF state eigenvalue equations 
for a dielectric slab cavity with $n_0=2$. (a) A QB state at 
$\tilde{k}_ma = 11.51917306 - 0.53647930i$. (b) A CF state at
$k_ma=11.55535804 - 0.53100809i$ with external wavevector 
$ka=11.51917306$. Since the eigenvalues
are almost the same the two 
solutions are almost identical within the cavity and show large 
amplification, but they differ qualitatively outside as the QB 
intensity grows exponentially while the CF intensity remains
constant 
as required for a true lasing solution.}
\label{figdislabQBCF}
\end{figure}

One can see that if $k$ is
chosen to be equal to $\tilde{q}_m$, the real part of the $m^{th}$ QB state
wavevector, then the two eigenvalue conditions are almost the same in
this vicinity.   Note however that for every choice of the external
wavevector $k$ there is an infinite
set of CF eigenvalues.  It is
shown in the Appendix that from that infinite set there is one 
$m^{th}$ CF state with a wavevector very close to that of
the corresponding QB state for $k=\tilde{q}_m$. Specifically, the difference between
the wavevector of the $m^{th}$ QB state
and the corresponding CF
state is given by
\begin{equation}
n_0(q_m -  \tilde{q}_m)a  = \frac{n_0 \kappa_0}{k};
\;\;\;\; n_0(\kappa_m - \kappa_0)a =
\frac{f(n_0)}{(ka)^2},
\end{equation}
where $f(n_0)$ is a number of
order unity depending on the index, $n_0$.  Hence this difference is
small in the semiclassical limit, $ka \gg 1 $.
Moreover both the CF and QB states are of the
form $A \sin(k_mx)$ within the dielectric, so if their complex
wavevectors are close, then the CF and QB states are almost identical
in the medium.
Thus we see that the standard cold-cavity resonances are almost equal
to the lasing mode,
but not exactly equal. Because the CF states are
sines of complex wavevector they will oscillate {\it and} increase
in amplitude within the medium (Fig.~\ref{figdislabQBCF}).

As noted in the general discussion, these CF
states are not orthogonal when integrated within the
medium.  It is therefore useful to define adjoint CF states which
satisfy the complex conjugate
differential equation, and with the {\it incoming }
boundary condition at infinity, $\partial_x \varphi_m(x = a) =
- i k \varphi_m(x =a )$.  It is easy to show that the
eigenvalues $\bar{k}_m$ and eigenfunctions $\bar{\varphi}_m$ of this 
problem are the complex conjugates of the CF eigenvalues $k_m$,
and eigenfunctions $\varphi_m$. In the Appendix it is confirmed that
these functions satisfy the orthogonality relation $\int_0^a dx 
\varphi_m (x) \bar{\varphi}^*_n (x) =
    \int_0^a dx \sin (n_0 k_m x)  \sin (n_0 k_n x )  =  \eta_m (\om)
\delta_{mn}$ and the
normalization constant $\eta_m (\om)$ is calculated.  The
biorthogonality of these functions follows from the
boundary
conditions at $x=a$.

\subsubsection{Cylindrical CF states}

The next case of interest is
the case of a circular (2D) or cylindrical (3D) dielectric resonator
of uniform
index $n_0$ and radius $R$.
The CF eigenvalues can be found by applying the conditions
Eqs.~(\ref{eqoutgoingCF}) and (\ref{eqcontBC0}) for the circle (2D) and
cylinder (3D, solutions uniform in z-direction).  Solutions
can be labeled by their good angular momentum index in the z-direction, M,
leading to a countably
infinite sequence of eigenvalues for each
value of M (which we will label by $m$).   The states take the form
\begin{equation}
\varphi_m^{(M)}(r,\phi,\om)=\begin{cases} J_M (n_0 k_m r) \ex{\pm i
M \phi}  & r < R \\
\frac{\besj{M}{n k_m R}}{\beshp{M}{k_m R}}\,\beshp{M}{k r}\, \ex{\pm iM
\phi} & r > R \end{cases}
\end{equation}
and the CF eigenvalue condition is found to be
\begin{equation}
\frac{\mbox{J}_M'(n_0 k_m R) \beshp{M}{k
R}}{\mbox{H}_M^{+\prime}(kR) \besj{M}{n_0 k_m R}} =
\frac{1}{n_0}\frac{k}{k_m}.
\end{equation}
Note that this equation differs from that defining the QB states for
this problem simply in that the corresponding QB eigenvalue condition
would have $k_m$ appearing in the arguments of $H_M^+$, the external
solutions, as well as in the arguments of $J_M$, the internal
solutions.

\begin{figure}[hbt]
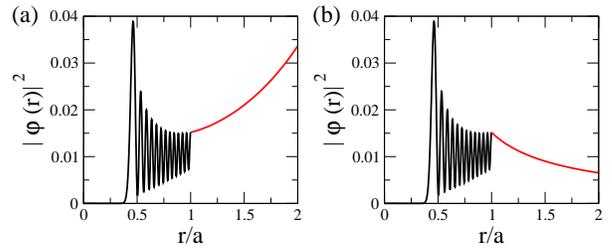

\includegraphics[clip,width=0.45\linewidth]{wfplotQB.eps}
\includegraphics[clip,width=0.45\linewidth]{wfplotCF.eps}
\caption{Solutions of the QB and CF state eigenvalue equations 
for a cylindrical dielectric resonator with $n_0=1.3$ and angular 
momentum quantum number $M=40$. (a) A QB state at $\tilde{k}_mR = 
71.17700357 - 0.74936509i$. (b) A CF state at
$k_mR=71.19539070 - 0.74853902i$ with external wavevector 
$kR=71.17700357$.  Again the two solutions are almost identical 
within the cavity; outside the QB solution grows while the CF 
solution decays as $1/r$ to conserve flux in 2D.  The amplification 
of the solutions within the cavity is less obvious since both 
solutions have the usual peak at the caustic at $r/a \approx 0.45$; 
the decay away from this peak
is actually slower going outwards then 
it would be for the cold cavity due to the amplification. }
\label{figcircleQBCF}
\end{figure}

As for the slab resonator discussed
above, if we choose $\om = c \tilde{q}_m$, the real part of a
specific QB state frequency, then we will find one CF eigenvalue
close to the complex QB state eigenvalue, as long as the QB state
itself has small imaginary frequency (high Q).  A comparison of two
such corresponding states is given in Fig.~\ref{figcircleQBCF}. 
Further analysis of
the 2D case is given in the Appendix.

\subsubsection{CF states for
general shapes}

The CF boundary conditions for a general dielectric body of arbitrary
shape are still easily formulated in terms of outgoing spherical
waves at infinity, since the object will appear point-like at
arbitrary large distances.  We simply require that the CF states
satisfy the wave equation with index $n_0$ within the medium and that
far away they have the form (in 2D)
\begin{equation}
\varphi_m (r \ra \infty,\phi) = f_m(\phi)\frac{1}{\sqrt{r}}\ex{ik r}.
\end{equation}
In order to satisfy this boundary condition it will be necessary to
solve the wave equation within the body by some method and continue
it sufficiently far outside to a sphere or circle enclosing the body
at which radius the interior solution can be connected to a
superposition of outgoing waves with the wave-vector $k$.  The
situation is simplified if the dielectric body has a smooth shape and
is not too far from spherical, in which case a Rayleigh (Bessel)
expansion can be implemented within the cavity and can be matched to
outgoing Hankel functions directly on the boundary specified by
$R(\phi)$:
\begin{equation}
\varphi_m (r,\phi)=\begin{cases} \sum_M \alpha_M \besj{M}{n k_m r}
\ex{iM \phi}\;\;\;  & r < R(\phi), \\
\sum_M \gamma_M \beshp{M}{kr} \ex{iM \phi}\;\;\; & r > R(\phi) \end{cases}
\end{equation}
The continuity boundary conditions on the boundary can be easily cast
into a secular equation and the eigenvalue condition replaced by a
singularity condition.  The method is then identical to that
described in reference \cite{Tureci05} for finding the QB
states of such an asymmetric resonant cavity (ARC) except that the
matrix $\cal{S}$ which appears from the matching conditions is
slightly different.  Two examples from implementing this method for
CF states are shown in Fig.~\ref{figbiARCwf1} below.

\begin{figure}[hbt]
\includegraphics[clip,width=0.45\linewidth]{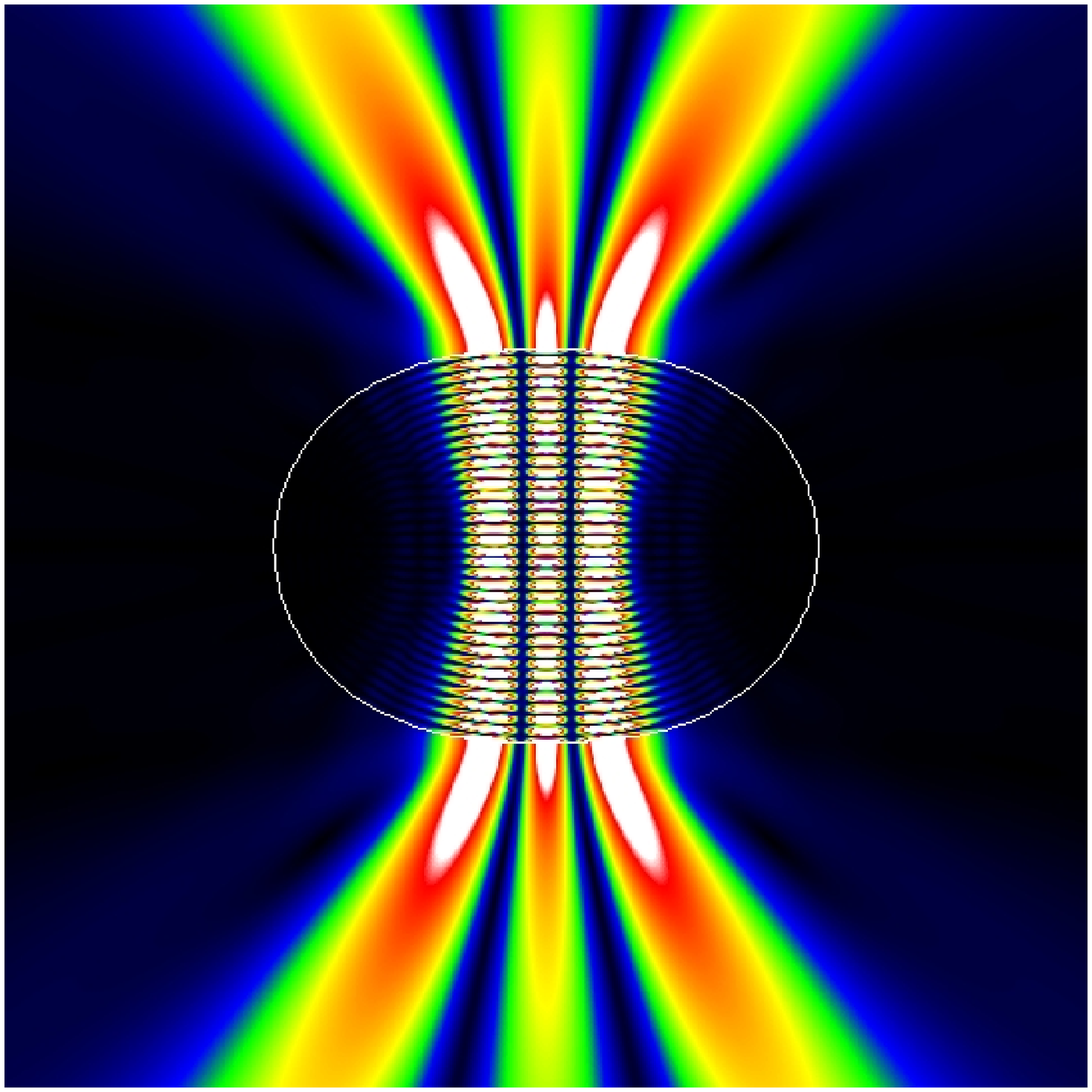}
\includegraphics[clip,width=0.45\linewidth]{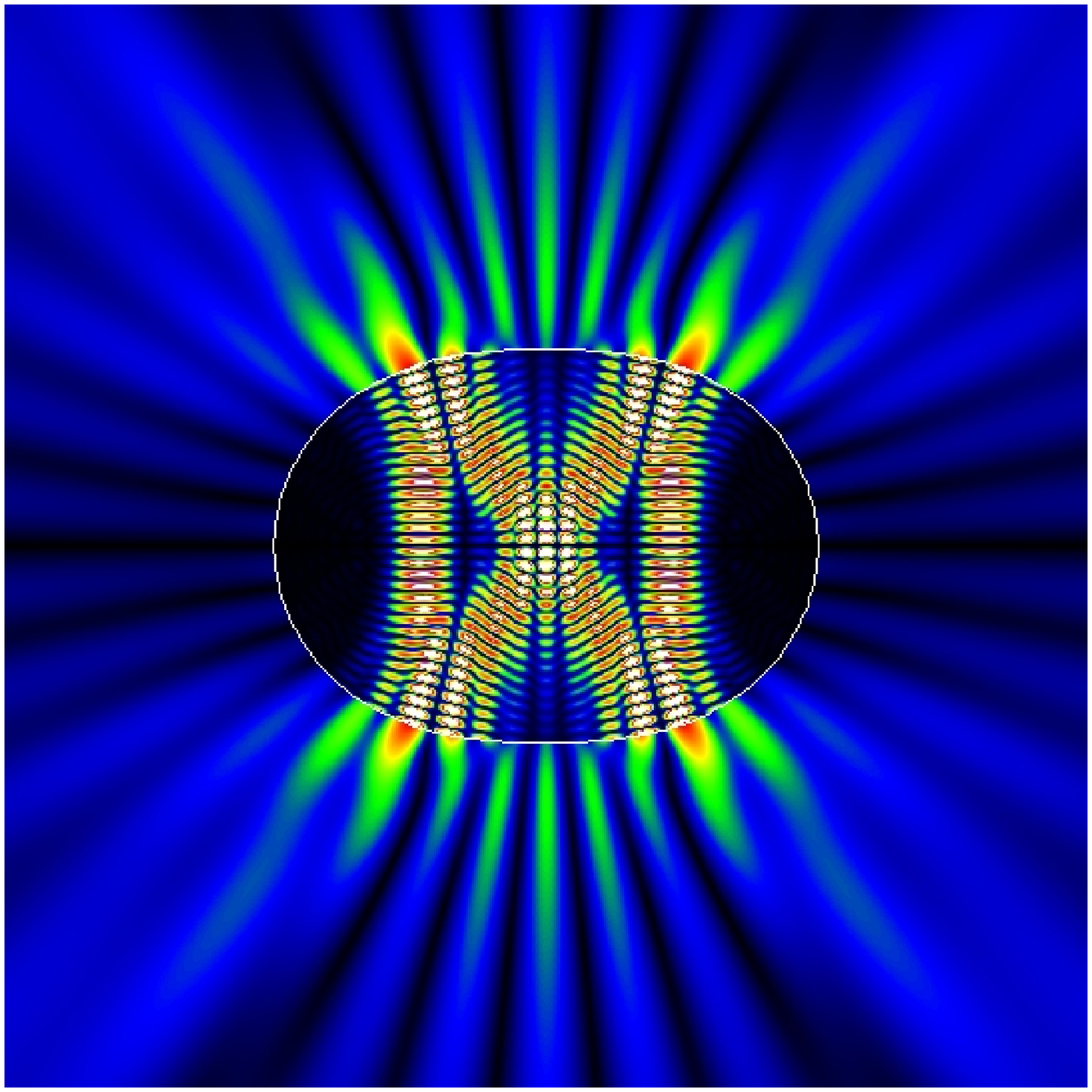}
\caption{Gray scale plots of CF solutions for a dielectric cavity 
with a quadrupolar deformation $R(\phi) = R_0 ( 1 + \eps\cos 2\phi )$ 
at $\eps=0.16$ and an index of refraction $n=3.3$
at external wavevector $kR_0=20.0$. (Left) A bouncing ball mode at 
$k_mR_0 =20.48417472 - 0.09782270i$ and (Right) a bowtie mode at $k_mR_0 
= 19.71417046 - 0.040257141i$.  Modes of the bowtie type were found 
to lase in the experiments of Ref. \cite{science98}.}
\label{figbiARCwf1}
\end{figure}

\section{Generalized multimode lasing equations}
\label{sectgenmmeqns}
\subsection{Non-linear multimode equations}

We now return to the formulation of the multi-mode lasing equations
within the single-pole approximation to the Green Function.
The outgoing lasing field  $e(x,\om)$ is only non-zero
at the lasing frequencies; we approximate the Green function in the
vicinity of $\om = \Om_\mu$ by
the term arising from the nearest pole,
i.e. $G(\bx,\bxp |\om =  \Om_\mu) \approx
( c^2 \varphi_{\mu}(\bx,\Om_\mu)\bar{\varphi}^*_{\mu}(\bxp,\Om_\mu))/( 2 n^2
\om_a (\Om_\mu -\om_{\mu}(\Om_\mu))$.
Equation (\ref{eqscint}) now takes the form:
\begin{widetext}
\begin{equation}
a_\mu \left( 1 - i\tilde{D}_0 \frac{1}{ (-\delta \om_\mu
+ i \kappa_\mu )}
\frac{\g}{(-i \Omu +\gamma_{\perp})} \int_{cavity}
d\bxp
\frac{ \bar{\varphi}^*_{\mu}(\bxp,\Om_\mu) \varphi_\mu
(\bxp,\Om_\mu)} {1 + \sum_\nu g(\Om_\nu) |a_\nu|^2 |\varphi_\nu
(\bxp,\Om_\nu) |^2} \right) = 0
\label{eqhsnl1}
\end{equation}
\end{widetext}
where $\tilde{D}_0$ is the scaled pumping rate given by
\begin{equation}
\tilde{D}_0 = \frac{2\pi \omega_a g^2}{\hbar n^2 \g} D_0
\end{equation}
We are looking for lasing solutions to these equations, i.e.
solutions where at least one $a_\mu \neq 0$
and positive.  Since $
a_\mu = 0, \forall_\mu$ is always a solution for the steady-state
equations, once we reach a pump level at which non-zero solutions
exist, we must check their stability.

This can only be done by
writing time-dependent equations for the CF state
amplitudes  $a_\mu
(t)$ (analogous to the standard modal expansions) and
determining the effect of a small deviation from the coexisting
solutions with $a_\mu = 0$ and $a_\mu \neq 0$. Previous work
based on the standard theory finds \cite{FuH91}
that whenever a solution with $a_\mu \neq 0$ exists, the $a_\mu = 0$
solutions are
unstable to small deviations, which then flow to the finite amplitude
solution with the largest number of lasing modes.  We assume
this property for the current work and will explore this stability
issue in subsequent work.
Ruling out the identically zero solutions, and defining $I_\nu \equiv
|a_\nu|^2$, the general non-linear multimode problem can be cast in
the form
\begin{equation}
       1  +          \frac{ i (-\delta \om_\mu
+ i \kappa_\mu )(-i \Omu +\gamma_{\perp})}{\tilde{D}_0 \gamma_{\perp}}
= F_\mu (I_\nu )
\label{eqhsnl2}
\end{equation}
where
\begin{equation}
F_\mu (I_\nu)
\equiv 1- \int_{cavity} d\bxp
\frac{ \bar{\varphi}^*_{\mu}(\bxp,\Om_\mu) \varphi_\mu
(\bxp,\Om_\mu)} {1 + \sum_\nu g(\Om_\nu) I_\nu |\varphi_\nu
(\bxp,\Om_\nu) |^2} ,
\label{eqfmuInu1}
\end{equation}
and we must look for the solution with the maximum number of non-zero
positive $I_\nu$.
Note that the CF states $\varphi_\mu
(\bx)$ in the definition of $F_\mu$ are
understood to be evaluated for $\om = \Om_\mu$ and normalized to
unity at this frequency.  We will refer to the full non-linear form of the
self-consistent equation as the non-linear Haken-Sauermann (NLHS)
equations; these equations can be treated numerically and we will
present results for the single and two-mode cases below.

\subsection{Generalized Haken-Sauermann equations}
To treat the multi-mode case more simply and to make contact with the standard
Haken-Sauermann treatment we now make the near threshold
approximation for  $F_\mu$ expanding the denominator of the integrand as
$1/(1+\sum_\nu
g(\Om_\nu)  | a_\nu \varphi_\nu (\bxp)|^2) \approx 1 - \sum_\nu
g(\Om_\nu)  | a_\nu \varphi_\nu (\bxp)|^2  $. We then arrive at the
generalized (linear) Haken-Sauermann (GHS) equations for the modal intensities:
\begin{equation}
1 + \frac{ i (-\delta \om_\mu
+ i \kappa_\mu )(-i \Omu +\gamma_{\perp})}{\tilde{D}_0 \gamma_{\perp}}
      = \sum_{\nu}
g(\Om_\nu) A_{\mu\nu} I_\nu.
\label{eqghs}
\end{equation}
where the overlap matrix $A_{\mu\nu}$ is given by
\begin{equation}
A_{\mu\nu} = \int d\bm{x}'  \, \bar{\varphi}_{\mu}^{*}(\bxp)
\varphi_\mu  (\bxp) | \varphi_{\nu}(\bxp)|^2 .
\end{equation}
In the limit in which $\gamma_{\perp} \gg
\Om_\mu$ and the frequency shift
$\delta \om_\mu \approx 0$ we recover the standard
form of the HS equation (Eq.~(\ref{eqdiagnin1}) above), with the
important difference that
the overlap matrix $A_{\mu \nu}$ is differently defined; it is complex and
involves the integral of three outgoing CF states and one
biorthogonal incoming state, whereas the usual HS equations
$A_{\mu \nu}$ involves the squared moduli of two closed cavity states
and is purely real.  This new form
allows us to determine iteratively
the lasing frequency shifts in the multi-mode case, whereas the
real approximation does not.  A second difference is that in the
usual HS equations the cavity widths $\kappa_\mu$  are put in by hand
at this point, as the closed cavity states have zero width.  In our
approach the widths appear automatically as the imaginary parts of
the complex wavevectors of the CF states.  Finally we note that in
this formulation the theory of interacting lasing modes
can be seen to have some similarity to the theory of interacting electrons in
quantum dots, as the overlap matrix $A_{\mu \nu}$ is similar to a mesoscopic
two-body matrix element.  When the resonator involved has complex
(e.g. wave-chaotic) modes
these matrix elements will fluctuate and
depend sensitively on external parameters such as shape
and pump
profile.  Concepts from random matrix theory and semiclassical
quantum mechanics may be
useful in analyzing these interactions; an example of this approach
using the usual near threshold HS equations
is the work of Misirpashaev and Beenakker
\cite{misirpashaevB98}.

\section{Solution of the generalized
HS equations}
\label{sectitsolgenhseq}
\subsection{Iterative method}
\label{subsectitmeth}

Both the non-linear form (\ref{eqscint}) and linear form
(\ref{eqghs})  of the generalized HS equations
require a
self-consistent iterative solution since the function $F_\mu (I_\nu)$
in its exact and linearized form
depends on the unknown lasing
frequencies $\Omu$ as well as the unknown intensities $I_\nu$.
Recall
that the quantities $\delta \om_\mu$ and $\kappa_\mu$ which
appear explicitly in the NLHS and GHS
equations are the  difference
between the lasing frequency $\Omu$
and the nearest CF state complex frequency $\om_\mu(\Om_\mu)$; thus
both of these quantities are
determined by the real lasing
frequencies.  Therefore an initial ansatz for the
lasing frequencies
is required in order to begin the root search to determine the $
\{I_\nu \}$.
Conceptually the linear and non-linear cases are
solved by similar iterative schemes, the only difference
being that
in the linear case (GHS) the root search becomes equivalent to
inverting a linear
system. In the linear case the solution must observe
the constraint that $I_\nu \geq 0$ while in the non-linear case the positive roots, if they exist, have to be 
determined (as $I_\nu = 0$ is always a solution).

At least two approaches are possible.  One can start with the $N$
quasi-bound state frequencies which have real parts within a
linewidth $\g$ of $\om_a$ and use $\re{\om_\mu^{qb}} \equiv
\Omega_\mu^{qb} $ as the initial iterate
for the real lasing frequencies.
A second approach is to linearize
the eigenvalue
equation around the atomic frequency $\om_a$, assuming that the
imaginary parts of the CF frequencies don't change rapidly so that
one can use the imaginary parts determined by solving
the CF eigenvalue condition at $\om_a$ as a starting point for the
iteration.  Each of these approaches has advantages and can
lead to a tractable scheme, as we will discuss below in the context
of specific examples. We will describe the first approach here.

We first separate the real and imaginary parts of Eq.~(\ref{eqghs})
\begin{eqnarray}
1-\frac{1}{D_0\g}\left[ \delta\om_\mu \Omu +  \kappa_\mu \g
\right] & = & \re{F_\mu} \label{eqFmunuR}\\
-\delta\om_\mu \g + \Omu  \kappa_\mu & = & D_0\g \im{F_\mu}
\label{eqFmunuI}
\end{eqnarray}
where we have dropped the tilde on $D_0$.  Our initial guess for the
lasing frequencies is $\Om_\mu^{(0)} = \Om_\mu ^{qb}$. The quantities
$\delta \om_\mu , \Omu, \kappa_\mu, F_\mu$ in Eq.~(\ref{eqFmunuR})
are all functions of the true lasing frequencies $\Omu$. However we
will approximate them by their values at the  $\Om_\mu^{(0)}$: let $
\delta \om_\mu^{(0)} = \re{\om_\mu(\Om_\mu^{(0)})} - \Om_\mu^{(0)}$,
$\kappa_\mu^{(0)} = -\im{\om_\mu(\Om_\mu^{(0)})}$,
$F_\mu = F_\mu
(\{ \Om_\nu^{(0)}\}) \equiv F_\mu^{(0)}$;
Eq.~\ref{eqFmunuR} then becomes
\begin{equation}
     1-\frac{1}{D_0\g}\left[ \delta\om_\mu^{(0)}
\Omu^{(0)} +  \kappa_\mu^{(0)} \g
\right]   = \re{F_\mu^{(0)}(\{I_\nu \})}.
\label{eqmodintit0}
\end{equation}
This equation can be solved for the modal intensities $I_\nu^{(0)}$
by varying the pump power and looking for positive roots for
different trial sets of lasing modes.  In practice one easily finds
the threshold for the first mode to lase and then increases the pump
in steps above this.

Having determined the intensities, Eq.~(\ref{eqFmunuR}) yields an
expression for the lasing frequencies by substituting $\Om_\mu =
\Omu^{(0)} + \delta \Om_\mu$, $\delta \om_\mu = \delta \om_\mu^{(0)}
- \delta \Om_\mu$, $\kappa_\mu =  \kappa_\mu^{(0)}$ and solving for
$\delta \Omu$
\begin{equation}
\delta \Om_\mu = \frac{\g(\delta \om_\mu^{(0)} +  D_0
\im{F_\mu^{(0)}})  - \kappa_\mu^{(0)} \Omu^{(0)}     }{\g + \kappa_\mu^{(0)} }.
\label{eqlasshiftit0}
\end{equation}
These equations yield the updated lasing frequencies $\Om_\mu^{(1)}$
which can be reinserted into the equations to obtain higher
iterations. We should note that Eqs.~(\ref{eqmodintit0}) and
(\ref{eqlasshiftit0}) are correct to O($\delta \Om_\mu$) and O($\delta
\Om_\mu^2$) respectively, and we have neglected terms of O($\partial
\om_\mu / \partial \om$) which can be shown to be small in the short 
wavelength limit. 
Thus Eqs.~(\ref{eqmodintit0}) and 
(\ref{eqlasshiftit0}) are
expected to converge rapidly with
iteration number.

The equation for the frequency shifts is a generalization of
well-known results for the single-mode case and the closed cavity.
First note that for the closed cavity the function $F_\mu$ is real
and doesn't enter the imaginary part of the HS equations; hence the
pump strength drops out.  In addition, in a standard treatment the
real frequency shift $\delta \om_\mu^{(0)}$ is assumed to be zero 
since one begins with the cold cavity
frequency, so one obtains:
\begin{equation}
\delta \Om_\mu
= \frac{ - \kappa_\mu^{(0)} \Omu^{(0)}     }{\g + \kappa_\mu^{(0)} }.
\end{equation}
This gives the well known result that when the cavity width is
large compared to the atomic relaxation rate one obtains $\delta
\Om_\mu = - \Omu^{(0)} \rightarrow \Omu = 0$ i.e. lasing at the
atomic frequency $\om_a$; when $ \kappa_\mu^{(0)} << \g$ (the cavity
mode is much narrower than the atomic linewidth) one obtains $\delta
\Om_\mu =  -(\kappa_\mu^{(0)}/\g) \Omu^{(0)} \rightarrow \Omu \approx
\Omu^{(0)} = \re{\om_\mu^{qb}}$, i.e. lasing at the cold cavity
frequency. Our generalization shows that  openness of the cavity
introduces the pump strength into these  equations and implies a
change of the lasing frequencies with pump strength.  This is a
prediction of our theory which can in principle be tested experimentally.

\subsection{Single mode solutions}
\label{subsectsm}

\subsubsection{Near-threshold behavior}

The simplest solution of Eqs.~(\ref{eqmodintit0}) and
(\ref{eqlasshiftit0}) is the case where the electric field is
oscillating at a single frequency $\Omu$. Then, we can solve
Eqs.~(\ref{eqmodintit0}) and (\ref{eqlasshiftit0}) straightforwardly
for $I_\mu$ and $\Om_\mu$:
\begin{equation}
I_\mu =  \frac{1}{g(\Omu^{(0)})A_{\mu\mu}^{(0)}}\left( 1 -
\frac{\kappa_\mu^{(0)}}{D_0} \right)
\end{equation}
With uniform pumping this equation implies that the first mode to
lase (and the only one in the single-mode approximation) is indeed
the mode with the highest Q, corresponding to the CF state with the
smallest value of its imaginary wavevector $\kappa_\mu$.
(Of course, in a realistic
scenario the laser typically will not remain single-mode for the whole range of
pump rates $D_0$ and other modes can begin to oscillate).
As noted above, unlike the closed cavity single-mode theory, for which the
lasing frequency is independent of pump strength, in this case we find
\begin{equation}
\Omu = \frac{\g \om_\mu^{(0)} + \g D_0 g(\Omu^{(0)})
\im{A_{\mu\mu}^{(0)}} I_\mu}{\g + \kappa_\mu^{(0)}}.
\label{eqsmlasfreq1}
\end{equation}
The first term in the numerator gives the ``center of mass" formula
for the lasing frequency discussed above, but the second term is
proportional to $I_\mu$ and demonstrates that there is a frequency
pulling or pushing effect even in the single-mode case which depends
on the pump strength and on the imaginary part of the inverse mode
volume, $A_{\mu \mu}^{-1}$.
The generalization to multi-mode lasing is
here trivial with the factor $\im{A_{\mu \mu}} I_\mu$ simply replaced
by $\sum_\nu \im{A_{\mu \nu}} I_\nu$ (cf. Eq.~(\ref{eqghs})).

\subsubsection{Far from threshold behavior}

To generalize the above results to arbitrary pump
rates we simply replace in Eqs.~(\ref{eqmodintit0}) and (\ref{eqlasshiftit0})
$F_\mu$ by its exact expression
\begin{equation}
F_\mu = 1 - \int_{cavity}
d\bxp
\frac{ \bar{\varphi}^*_{\mu}(\bxp) \varphi_\mu (\bxp)} {1 +
g(\Om_\mu) |\varphi_\mu (\bxp) |^2 I_\mu},
\end{equation}
where it is understood that the CF states are calculated at the
iteration frequency $\Om_\mu=\Om_\mu^{(i)}$.

\begin{figure}[hbt]
\includegraphics[clip,width=\linewidth]{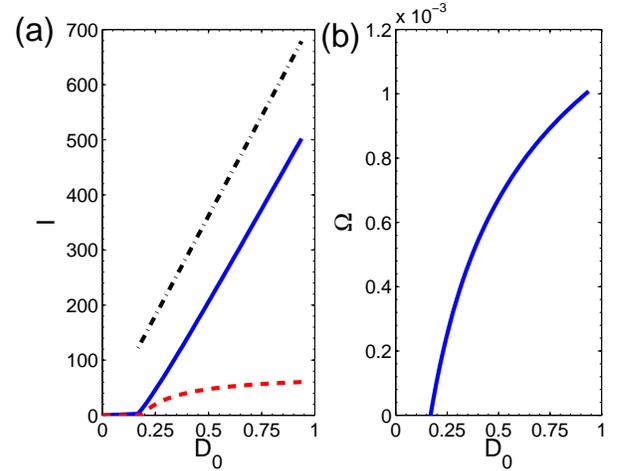}
\caption{Single-mode single-pole solutions for the modal intensity 
vs. pump strength $D_0$ for
a circular dielectric
resonator with an index of refraction $n=2.0$ and gain-center at
$\om_a=50.0$. (a) Single-mode solution for the intensity $I$ of the
mode with an angular momentum quantum number $M=40$ lying closest to
the gain center. Dashed line: near-threshold solution, solid line: exact
solution for modal intensities $I$ and dot-dashed line:  far-from
threshold approximation discussed in the text (plotted only for $I
>120$). (b) Lasing frequency $\Om$ as
a function of the pump strength $D_0$. Note that $\Om$ is measured
from its value at the lasing threshold given by $\Om_0 = 49.6668$ and 
the shift is
towards the gain center frequency.}
\label{figsmsp1}
\end{figure}

In Fig.~\ref{figsmsp1} we show an example of the proposed iteration 
scheme for a dielectric
resonator of circular cross-section. There are several observations
we can make here. First of all, the intensity calculated from the
near-threshold theory greatly underestimates the actual intensity at
higher pump strengths $D_0$. As we shall further discuss below in
Section~\ref{subsect2m}, this will have important consequences for 
mode competition in
the multimode regime as the non-linear thresholds will be substantially
different from those predicted by the near-threshold approximation.
Second, as seen in Fig.~\ref{figsmsp1}(b) we now get a frequency shift that is
power-dependent, with a non-linear dependence on the lasing
intensities.  This dependence is stronger the leakier the lasing
mode. Third, the exact formula leads to an approximately linear 
dependence of the modal
intensity $I_\mu$ on $D_0$.  This linear dependence is termed
"saturation'' in laser texts because it is much slower than the near 
threshold rise, but it is
still a much stronger dependence than predicted by the near threshold
HS theory.   One can obtain a naive approximation for $I_\mu$ by
replacing $ 1  + g(\Om_\mu) |\varphi_\mu (\bxp) |^2 I_\mu$ in the
denominator with $g(\Om_\mu) |\varphi_\mu (\bxp) |^2 I_\mu$, then
\begin{equation}
I_\mu = \left( \frac{\re{\Phi_\mu}}{g(\Om_\mu)\kappa_\mu} \right) D_0
\end{equation}
where
\begin{equation}
\Phi_\mu = \int_{cavity} d\bm{x}' \, \ex{2i\phi_\mu(\bm{x}')}
\end{equation}
where $\phi_\mu(\bm{x}') = -i \text{Arg} [\varphi_\mu(\bm{x}')]$. For
comparison, we plot the result of this expression in
Fig.~\ref{figsmsp1}(b) (dot-dashed line). We see that this formula is
qualitatively correct but overestimates the modal intensity.

\subsection{Two-mode solutions}
\label{subsect2m}

\begin{figure}[hbt]
\includegraphics[clip,width=\linewidth]{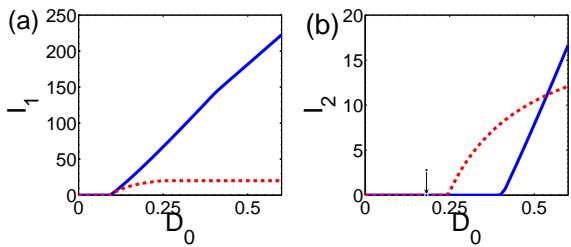}
\caption{A single-pole two-mode solution for the modal intensities as a function of pump strength $D_0$ for a circular dielectric
resonator with an index of refraction $n=2.0$ and gain-center at
$\om_a=20.0$. We have chosen $\gamma_\perp=50$ using effectively a 
flat gain curve. The two modes have the angular momentum quantum 
numbers $M_1=20$ and (b) $M_2=18$ respectively. The modal intensities 
$I_1$ and $I_2$ are given in (a) and (b). Solid lines denotes the 
exact solution, while we plot the near threshold approximate solution 
in dashed lines. The non-interacting thresholds are given by 
$D_{th}^{(1)}=0.1069$ and $D_{th}^{(2)}=0.1767$ (marked with vertical 
arrow in (b)).  By definition modal interactions cannot change the 
first threshold, which is thus the same in the near-threshold and 
exact solution.  However interactions dramatically modify subsequent 
thresholds; in this case the second
threshold is increased to 
$D_{th}^{(2)}=0.2464$ in the near threshold approximation, but 
actually should be increased much more,  to $D_{th}^{(2)}=0.4139$, 
based on the exact solution.  This indicates that even
taking modal 
interactions into account through the near-threshold approximation is 
inadequate and likely greatly overestimates the number of lasing 
modes at a given pump power.}
\label{figtmsp1}
\end{figure}

\begin{figure}[hbt]
\includegraphics[clip,width=\linewidth]{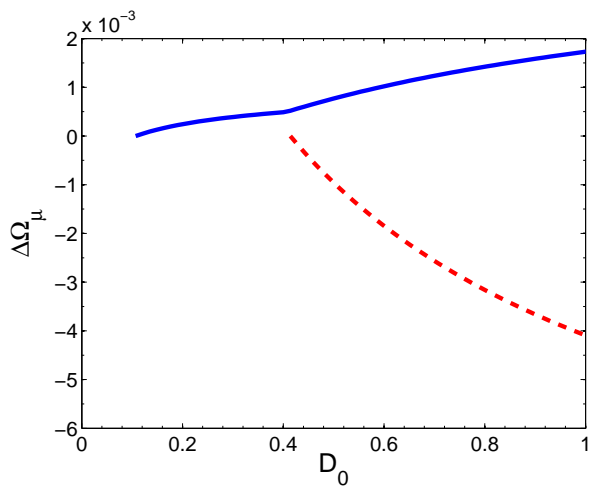}
\caption{The corresponding laser frequency variation with pump power 
for the example in Fig.~\ref{figtmsp1}  as a function of  pump 
strength $D_0$, for the exact solution. The frequencies of the two 
modes are measured with respect to their values at their respective 
lasing thresholds given by $\Om_1 = 19.8584$ (full line) and $\Om_2 = 
20.3764$ (dashed line).  Note that since the gain center is $\omega_a 
= 20.0
$ both modes are pulled towards the gain center.}
\label{figtmsp2}
\end{figure}

The numerical effort required to solve the linear and non-linear
generalized HS equations for
a non-trivial case, e.g. a dielectric
cavity with fifty modes underneath the gain curve, is
substantial.
Moreover the interesting questions require varying the
pump strength and the shape of the cavity.
In earlier work we have
done extensive calculations along these lines using the near
threshold HS equations
for the closed cavity and then using
reasonable but uncontrolled approximations to evaluate the
modal
intensities outside the cavity \cite{tureci05b}.  The current
generalized formalism gives us a method to
evaluate emission
directionality and output intensities correctly and without any
approximations beyond
the usual ones allowing a multi-mode solution.
It also presents a formulation of the lasing equations
valid away
from the threshold, using the general function, $F(I_\nu)$, defined
in Eq.~(\ref{eqfmuInu1}) above.
A striking shortcoming of the near
threshold approximation that we found in our earlier work \cite{tureci05b}
is
that it appears to overestimate greatly the number of lasing modes
compared to the experimental
observations \cite{science98}.  Here we
will not present realistic calculations on complex cavities using
our
generalized formalism; we defer such calculations to future work.
Instead we present a simple comparison
of the near-threshold and
non-linear lasing solutions for the two-mode case, which confirms that
the widely-used near threshold approximation can
introduce a large quantitative error in the lasing thresholds and
hence overestimates greatly the number of lasing modes for a given
pump power in the complex
dielectric cavity lasers we are studying.
We solve Eq.~(\ref{eqhsnl2}) for two interacting modes of the
dielectric cylinder
(other modes which might also lase are
neglected).  In Fig.~\ref{figtmsp1} and  \ref{figtmsp2} we show the 
results.  For mode 1, with the
lower threshold,  the near-threshold and non-linear
calculation give the same threshold as they must ($D_0 \approx 0.11$), and then
deviate substantially well above threshold as we already saw in the
previous section.  Mode 2, which would have a threshold
of $D_0 \approx 0.18$ without mode competition has a threshold of $D_0 \approx
0.25$ in the near threshold approximation,
including the effects of
mode competition, but has a correct threshold, taking non-linear
effects into account, of
$D_0 \approx 0.41$.  Hence we see that in
the large interval $0.25 < D_0 < 0.41$ the near-threshold theory
predicts
that two modes would be lasing and with comparable
intensity, while the full theory predicts only one lasing
mode.
Moreover, because of this delay, when Mode 2 does begin to
lase, its intensity lags mode 1 by an order of magnitude
in contrast
to the prediction of the near threshold analysis.  We expect these
discrepancies to be enhanced if there
are many competing modes and
many fewer modes to be lasing at a given pump power than predicted
by
the near-threshold theory.  These issues will be explored in
detail in future work.

\section{Summary and Conclusions}
\label{sectconcl}

We have reformulated
semiclassical lasing theory to treat open cavities, with particular
attention to complex and random lasers, for which the output power
and directional emission patterns are not trivially found from
knowledge of the internal modal intensities.  We begin from the
assumption that a steady-state multi-periodic lasing solution
exists and is stable. A key idea is to formulate the theory in terms
of the self-consistent linear response to the polarization described
by an outgoing Green function.  This yields a set of self-consistent
integral equations for the lasing modes which can be solved by
various means.  The approach which is closest to conventional
multi-mode lasing theory is to write a spectral representation of
this Green function in terms of outgoing states with constant flux.
For high Q cavities these states are similar but distinct from the
usual resonances or quasi-bound states; they satisfy useful
biorthogonality relations with adjoint functions.  The Green function
is then approximated by a single-pole near each lasing frequency
implying the lasing mode is just an intensity $I_\mu$ times a single
CF state.  General and near-threshold equations then follow for these
intensities $I_\mu$ and for the lasing frequencies $\Om_\mu$.  The
near threshold equations are generalizations of the well known
Haken-Sauermann equations derived for a closed cavity.  These
equations can account for the effects of mode competition and spatial
hole-burning which are important but difficult to treat in complex
cavity lasers.  An iterative scheme was described for solving these
equations and some illustrative results were given for the single-mode
and two-mode case.  The one and two-mode solutions indicate that the
near-threshold
approximation substantially overestimates the number
lasing modes well above
threshold and underestimates their output
power.

In our view this approach clarifies the
longstanding question of how to treat rigorously open cavities within
semiclassical lasing theory.  It now remains to apply this formalism
to cases of interest: dielectric micro-cavity lasers of interesting
shape, cavities with fully chaotic ray motion, and lasing from random
media.

\appendix
\section{\label{sectappCFprop}Properties of the constant flux basis}

In this Appendix we will explore the constant flux (CF) states, as
defined in Sec.~\ref{sect1dCF}.  We investigate the
analytically tractable cases of the 1D dielectric slab laser and the
2D cylindrical laser, and argue that the
qualitative features of the solutions for these simple geometries
should also hold for more general shapes.
Here we focus on three
properties of the CF basis.  First, that for each high-Q QB state of
the cavity there exists
a single CF state with similar real and
imaginary part and hence very similar behavior within the cavity
(where they satisfy the same differential equation with just a
slightly shifted eigenvalue).  Second, we prove that all the CF
states have eigenvalues with negative imaginary parts, corresponding
to amplification of the wave within the cavity.
Third, we confirm
explicitly the biorthogonality of the two (outgoing and incoming) CF
bases within the cavity
for a general geometry.

\subsection{Dielectric slab cavity}
\label{sectappdislab}

As stated in Sec.~\ref{sect1dCF}, for a semi-infinite dielectric
``slab'' resonator
the QB and CF eigenvalue equations are:
\begin{equation}
\tan{(n\tilde{k}_m a)}=-in
\label{eqqb_ev}
\end{equation}
and
\begin{equation}
\tan{(nk_m a)}=-in \left( \frac{q_m-i\kappa_m}{k} \right)
\label{eqcf_ev}
\end{equation}
Where $\tilde{k}_m \equiv \tilde{q}_m - i\tilde{\kappa}_m$ and ${k_m} \equiv q_m - i\kappa_m$ are the 
complex QB and CF
wavenumbers, $a$ is the length of the slab, $n$ is the (possibly
complex) index of
refraction, and $k$ is the wavenumber corresponding to the
external' Fourier frequency.  As in the main text we use
tildes to denote quantities associated with the QB states, but we 
drop the subscript $0$ on the
index of refraction.  We also define 
the dimensionless quantities $x_m=n q_m a$,
$y_m =-n \kappa_m a$, and $z=n k a$.
Using trigonometric identities, we can expand
both equations into real and imaginary parts:
\begin{equation}
\frac{\cot{\xt} (1-\tanh^2{\yt})}{\cot^2{\xt}+\tanh^2{\yt}}=0
\label{eqqb_ev1}
\end{equation}
\begin{equation}
\frac{\tanh{\yt}(\cot^2{\xt}+1)}{\cot^2{\xt}+\tanh^2{\yt}}=-n
\label{eqqb_ev2}
\end{equation}
and
\begin{equation}
\frac{\cot{\x} (1-\tanh^2{\y})}{\cot^2{\x}+\tanh^2{\y}}=\frac{n\y}{\z}
\label{eqcf_ev1}
\end{equation}
\begin{equation}
\frac{\tanh{\y}(\cot^2{\x}+1)}{\cot^2{\x}+\tanh^2{\y}}=-\frac{n\x}{\z}
\label{eqcf_ev2}
\end{equation}
In the method used to solve the multimode laser equations presented
in this paper, the Green's function is
approximated by a single pole, i.e. by a single CF state in the
spectral representation of Eq.~(\ref{eqspecrep2}).  We are
therefore especially interested in finding  solutions to the CF
eigenvalue equation with real part close to the
wavevector, k, appearing as a parameter in the CF eigenvalue equation.
Furthermore, since we expect the real part of the QB mode frequencies to provide a first approximation to the 
actual lasing
frequencies, we look for solutions to the CF eigenvalue equations
with the external Fourier frequency set equal
to the real part of a quasibound mode frequency.  Equations
(\ref{eqqb_ev1})-(\ref{eqcf_ev2})
suggest that, for each QB mode with a sufficiently
short wavelength, a corresponding CF solution exists with
approximately the same eigenvalue, since in this limit
    $x=n\tilde{k}_ma$ is large.  In
that case the right hand side of (\ref{eqcf_ev1}) is small so that
(\ref{eqcf_ev1}) is nearly
identical to (\ref{eqqb_ev1}), and for $q_m \approx x$
equation (\ref{eqcf_ev2}) approaches (\ref{eqqb_ev2}).

\begin{figure}
\psfrag{A}{$\frac{\tilde{q}_m-q_m}{\tilde{q}_m}$}
\psfrag{B}{$\tilde{q}_m$}
\psfrag{C}{$\frac{\tilde{\kappa}_m-\kappa_m}{\tilde{\kappa}_m}$}
\psfrag{D}{$\ln{\tilde{\kappa}_m}$}
\psfrag{E}{$\tilde{q}_m$}
\includegraphics[width=\linewidth]{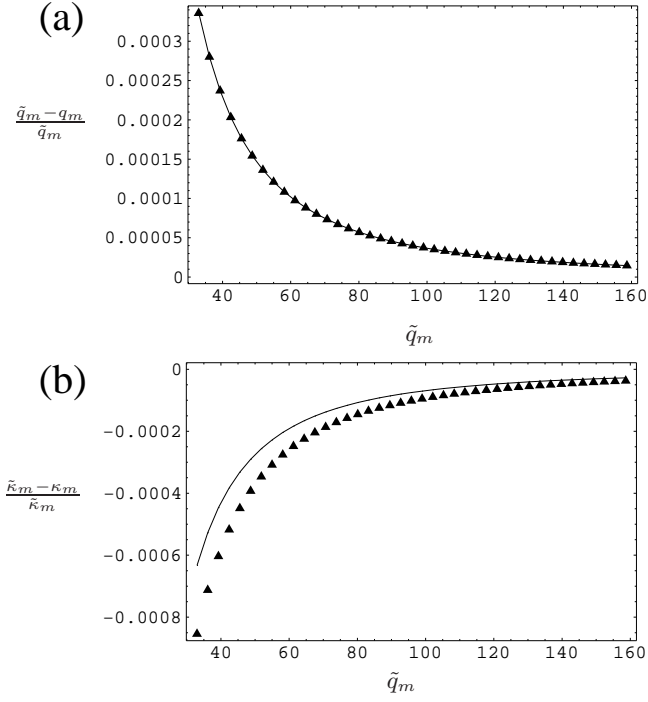}
\caption{The difference between the CF and QB eigenvalues for a 1D
dielectric slab resonator, as a function of the real
part of the QB eigenvalue.  (a) The fractional difference between the real part of the
eigenvalues, and (b) the fractional difference between the complex parts.  In both plots, the triangles
represent the actual difference, found numerically, which is
compared to the linear approximation, Eq.~(\ref{eqkappaslab}), 
derived in the text and represented by the solid 
curves.}
\label{fig_cf_qb_slab}
\end{figure}

The solution to Eqs. (\ref{eqqb_ev1}) and (\ref{eqqb_ev2}) was
given in the text by
equation (\ref{eq1dslabquant1}), and equations (\ref{eqcf_ev1}) and
(\ref{eqcf_ev2})
can be solved for $\tanh{\y}$ and $\cot{\x}$:
\begin{widetext}
\begin{equation}
\tanh{\y}=\frac{-(\cot^2{\x}+1) \pm \sqrt{(\cot^2{\x}+1)^2-4(n\x /
\z)^2\cot^2{\x}}}{2n\x/\z}
\label{eqcfsol1}
\end{equation}
\begin{equation}
\cot{\x}=\frac{(1-\tanh^2{\y}) \pm
\sqrt{(1-\tanh^2{\y})^2-4(n\y/\z)^2\tanh^2{\y}}}{2n\y/\z}
\label{eqcfsol2}
\end{equation}
\end{widetext}
Note that in order to self-consistently choose solutions to
(\ref{eqcfsol1}) and
(\ref{eqcfsol2}) which are close to QB solutions, we must
select the minus signs in both equations.
We now expand (\ref{eqcf_ev1}) in the small difference $\delx = q_{cf}-q_{qb}$:
\begin{equation}
\frac{\delta \x (1-1/n^2)}{1/n^2} \approx -\frac{n}{\z} \y
\end{equation}
Assuming that $\delta \y$ is, at most, first order in the small parameter
$\z^{-1}$, we can replace $\y$ with $\yt$, leading to:
\begin{equation}
\delta \x (n^2-1)  \approx  -\frac{n}{\z}\yt
\end{equation}
Or, using the expression for $\kappa_0$ from Eq. 
(\ref{eq1dslabquant1}) and the definitions of $\x$ and $z$:
\begin{equation}
\delta q_m \approx \frac{\ln{[\frac{n+1}{n-1}]}}{2(n^2-1)}\frac{1}{n a^2 k}
\label{eqdq_slab}
\end{equation}
We see the difference between real part of the CF resonance and the
external frequency is small when (1) the
external frequency is set near the real part of a QB resonance and
(2) the semiclassical parameter $\z^{-1}$ is
small.
To find an approximation for $\kappa_m$, as well as to justify the
approximation used in deriving (\ref{eqdq_slab}) that $\y \approx
\yt$, we next expand (\ref{eqcfsol1}) to second order in $\delx^{-1}$ and
$x^{-1}$.  After some manipulation we
obtain:
\begin{align}
\delta \kappa_m & \approx -\frac{1}{2(n^2-1)^2}\ln{[\frac{n+1}{n-1}]} \left(
1+\frac{\ln{[\frac{n+1}{n-1}]}}{2}
\right)\frac{1}{n a^3 k^2} \nonumber \\ & = \frac{f(n)}{(ka)^2} \frac{1}{a}
\label{eqkappaslab}
\end{align}
Which is the result quoted in Section~\ref{sect1dCF}.  Thus we see that $\delx$ is first
order and $\dely$ is second order in $z^{-1}$.
Fig. (\ref{fig_cf_qb_slab}) shows that
the approximation used here agrees well with the numerically
evaluated results for the
difference between the QB and the nearest
CF wavevector.

\subsection{Cylindrical dielectric cavity}
\label{sectdicyl}

It is well known that for a cylindrical cavity, there exist
whispering gallery mode
solutions to the QB eigenvalue equation that have very
small widths, $\tilde{\kappa}_m$.  Moreover, it is these modes which 
will dominate
lasing emission from dielectric
cylinder lasers. In this section we
show that there exist corresponding solutions to the CF eigenvalue
equations
close to these QB solutions.  Specifically, the difference between
the CF and QB lifetimes is proportional to a
small factor times the QB lifetimes; hence these CF and QB states
have almost identical lifetimes.

The QB eigenvalue equation for a cylindrical resonator is:
\begin{equation}
\frac{d}{dr}J_M(n\tilde{k}_m R)/J_M(n\tilde{k}_m
R)=\frac{d}{dr}H^{(+)}_M(\tilde{k}_m R)/H^{(+)}_M(\tilde{k}_m R)
\label{eqqbcyl_ev}
\end{equation}
Within the regime of interest where the modes are well trapped (and
the index of refraction is not too small), we
can approximate the Hankel and Bessel functions appearing in this
equation by the Debye approximations \cite{abramovitz}.  Define
$\tanh{\alpha}=\sqrt{1-(\tilde{k}_mR/M)^2}$ and
$\tan{\beta}=\sqrt{(n\tilde{k}_mR/M)^2-1}$.  In terms of these
variables, the
Debye approximations are:
\begin{equation}
J_M(k_mR)=\frac{e^{M(\tanh{\alpha}-\alpha)}}{\sqrt{2\pi M \tanh{\alpha}}}
\label{eqdeb1}
\end{equation}
\begin{equation}
Y_M(k_mR)=-\frac{e^{-M(\tanh{\alpha}-\alpha)}}{\sqrt{\frac{1}{2}\pi M
\tanh{\alpha}}}
\label{eqdeb2}
\end{equation}
\begin{equation}
J_M(nk_mR)=\frac{\cos{(M\tan{\beta}-M\beta-\frac{\pi}{4})}}{\sqrt{\frac{1}{2}\pi M \tanh{\alpha}}}
\label{eqdeb3}
\end{equation}
\begin{equation}
Y_M(nk_mR)=\frac{\sin{(M\tan{\beta}-M\beta-\frac{\pi}{4})}}{\sqrt{\frac{1}{2}\pi M \tanh{\alpha}}}
\label{eqdeb4}
\end{equation}
Note that in the regime in which these approximations are valid, we
can ignore the r-dependence outside the
exponent when we evaluate its derivatives, so that the eigenvalue
equation is approximately:

\begin{equation}
\tan{\beta}\tan{\lambda} \approx
\tanh{\alpha}\frac{e^{\mu}+2ie^{-\mu}}{e^{\mu}-2ie^{-\mu}}
\end{equation}
where we have defined $\lambda=M(\tan{\beta}-\beta)-\frac{\pi}{4}$
and $\mu=M(\tanh{\alpha}-\alpha)$.
By expanding in the small quantity $\tilde{\kappa}_m$ and neglecting
$e^{\mu}$ compared to $e^{-\mu}$ (which is valid
for well trapped modes), we can derive an approximate expression for
$\tilde{\kappa}_m$:
\begin{equation}
\tilde{\kappa}_m \approx \frac{M\tanh{\alpha}}{n^2 a^2 
\tilde{q}_m(n^2-1)}e^{2\mu}.
\label{eqqb_deb}
\end{equation}
A similar, but not identical, result was obtained by N\"ockel
\cite{noeckel_thesis}, where it was pointed out that the small
factor $e^{2\mu}$ could be interpreted as the exponential decay
factor due to tunneling.

\begin{figure}
\psfrag{A}{$\frac{\tilde{q}_m - q_m}{\tilde{q}_m}$}
\psfrag{B}{$\tilde{q}_m$}
\psfrag{C}{$\frac{\tilde{\kappa}_m-\kappa_m}{\tilde{\kappa}_m}$}
\psfrag{D}{$\ln{\tilde{\kappa}_m}$}
\psfrag{E}{$\tilde{q}_m$}
\psfrag{F}{$\frac{\kappa_m-\tilde{\kappa}_m}{\tilde{\kappa}_m}$}
\includegraphics[width=\linewidth]{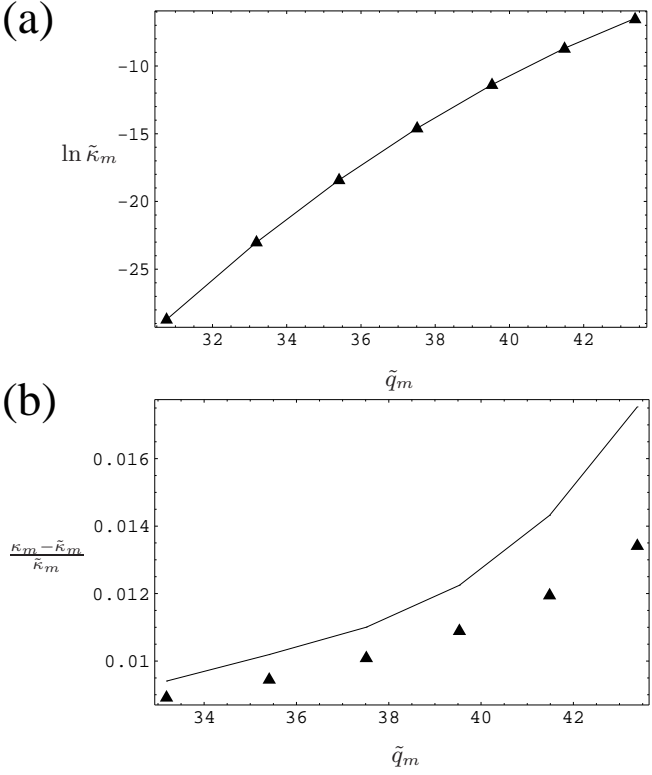}
\caption{(a) The log of the imaginary part of the QB resonance eigenvalue
for a cylindrical resonator with $n_0 = 2$, plotted against the real 
part of the eigenvalue; triangles
represent numerical results compared to 
the analytic approximation of Eq.~(\ref{eqqb_deb}). 
(b) The fractional difference between the CF and QB
lifetimes versus the real part of the QB
eigenvalue for the same resonator.  Stars are numerical results, 
solid line the 
approximation of Eq.~(\ref{eqdeltakmcyl}).  We have 
plotted the sequence corresponding to 
angular momentum $M=50$. }
\label{fig_kappaqb_dkappa}
\end{figure}

The eigenvalue equation for CF modes in a cylindrical resonator has
the same form as Eq. (\ref{eqqbcyl_ev}) above for QB
states except that the functions $\alpha$, $\beta$, $\lambda$ and
$\mu$ appearing in the CF equation are
evaluated at different arguments.  Explicitly, $\lambda$ and $\beta$
are evaluated at
$k_mR=\tilde{k}_mR+\delta(k_mR)$, whereas $\mu$ and $\alpha$ are
evaluated at the external Fourier frequency,
which we have set equal to $\tilde{q}_m$.  We can then expand in the
small quantities $\delta(k_mR)$ and
$\tilde{\kappa}_m$.  To first order, the imaginary part of this
equation yields:
\begin{equation}
\delta(\kappa_m) \approx 
\frac{\tilde{\kappa}_m}{M(n^2-1)\tanh{\alpha}},
\label{eqdeltakmcyl}
\end{equation}
where it is understood that $\tanh{\alpha}$ is evaluated at 
$\tanh{\alpha}=\sqrt{1-(\frac{\tilde{k}_mR}{M})^2}$
Note that in the semiclassical regime whispering gallery modes have a
high M value, so the difference between the
lifetimes for the CF and QB mode is a small factor times the already
small QB lifetime $\tilde{\kappa}_m$.

\subsection{General argument that $Im(k_m)=-\kappa_m <0$}
\label{sectimagneg}

We were able to show above explicitly that for the CF eigenvalue
closest to the corresponding
QB state eigenvalue, $\kappa_m >0$ for
both the 1D slab laser and the 2D cylindrical laser.
Here we show that this result holds generally for all CF eigenvalues
and for arbitrary geometries.
This is important because the exact
CF Green function involves all CF eigenvalues and it would
be
unphysical for it to have poles in the positive imaginary plane.

The
equation satisfied by the CF
states is:
\begin{equation}
\nabla^2\varphi(\bm{x})=\left\{ \begin{array}{ll}
				n^2k_m^2\varphi(\bm{x}) & 
\mbox{(inside)} \\
				k^2\varphi(\bm{x})   & \mbox{(outside)}
				\end{array}
				\right.
\end{equation}
We define a vector flux as
$\bm{f}=\frac{1}{2i}(\varphi^*\bm{\nabla}\varphi-\varphi
\bm{\nabla}\varphi^*)$.
The outgoing wave boundary conditions, as expressed in the main text
by equation (\ref{eqoutgoingCF}), imply that if we integrate
the radial component of $\bm{f}$ over increasingly larger spheres S
(or circles in the 2D case) with origin
within the cavity, the value of this integral approaches a positive
constant proportional to $k$:
\begin{equation}
\int_S \bm{f} \cdot \bm{dA} \longrightarrow k \int |\varphi(\phi)|^2 d\phi
\label{eqflux_int}
\end{equation}
Note that $\bm{\nabla} \cdot \bm{f}$ vanishes outside the cavity,
whereas inside it is equal to
$-n^2\varphi \varphi^* Im(k_m^2)=2n^2|\varphi|^2 q_m \kappa_m$.  By
using the divergence theorem and the CF equation we can also write 
this integral
as:
\begin{equation}
\int_S \bm{f} \cdot \bm{dA}=\int_{Int(S)}\bm{\nabla} 
\cdot \bm{f}
dV=2n^2q_m \kappa_m \int_{cavity}
n^2|\varphi|^2d\bm{x}
\label{eqflux_tot}
\end{equation}
Comparing equations (\ref{eqflux_int}) and (\ref{eqflux_tot}) we can
see that in order to satisfy
(\ref{eqflux_int}), which follows from the outgoing wave boundary
condition, we
must have $\kappa_m >0$.

\subsection{Biorthogonality of incoming
and outgoing CF states}
\label{sectbiorthCFARC}

Consider a solution to the CF eigenvalue equation $\varphi_m$ with
eigenvalue $k_m$, along with an adjoint solution $\overline{\varphi}
_n$ with eigenvalue $k_n ^*$.  Using the fact that these two
functions satisfy the
eigenvalue equations (\ref{eqcfgenev}) and (\ref{eqcfadjointgen}),
respectively, we have:
\begin{align}
n_0 ^2(k_n^2-k_m^2) & \int_{cavity} d\bx \, \overline{\varphi}^*_n \varphi_m
\nonumber \\ 
& = \int_{cavity} d\bx\,
\left(\overline{\varphi}^*_n \nabla^2 \varphi_m-\varphi_m \nabla^2
\overline{\varphi}^*_n
\right) \nonumber \\ & = \int_{boundary}(\overline{\varphi}^*_n \bm{\nabla}
{\varphi_m}-\varphi_m \bm{\nabla} \overline{\varphi}^*_n) \cdot
d\bm{A}
\label{eqcf_int}
\end{align}
Where the last integral is over the surface of the cavity boundary.
Outside the cavity we have
$\bm{\nabla}\cdot (\overline{\varphi}^*_n \nabla
{\varphi_m}-\varphi_m \nabla \overline{\varphi}^*_n)=0$, so we
can in fact replace the integral over the cavity boundary with an
integral over a large sphere $S$ inclosing
the cavity.  We can then invoke the outgoing/incoming wave boundary
conditions to evaluate the
surface integral explicitly:
\begin{align}
\int_S (\overline{\varphi}^*_n & \bm{\nabla} {\varphi_m}-\varphi_m 
\bm{\nabla}
\overline{\varphi}^*_n) \cdot
d\bm{A} & \nonumber \\ & \longrightarrow
\int_S (ik \overline{\varphi}^*_n {\varphi_m}-ik \varphi_m
\overline{\varphi}^*_n) \cdot
d\bm{A}=0
\label{eqcf_int1}
\end{align}
For the case $k_n \neq k_m$, equations (\ref{eqcf_int}) and
(\ref{eqcf_int1}) therefore imply:
\begin{equation}
\int_{cavity}d^3x \overline{\varphi}^*_m \varphi_n=0
\end{equation}
It should be noted that, although the CF functions are defined to exist both
inside and outside the cavity, the inner product is defined by an
integral over the cavity interior only.  The
biorthogonality of the CF states is therefore determined completely
by their behavior within the cavity and at
its boundary; we merely use the behavior of the CF functions at
infinity to show that the surface term in
equation (\ref{eqcf_int}) vanishes whenever the CF states satisfy the
outgoing/incoming boundary conditions.

We can check the biorthogonality of the CF states explicitly for the
case of the dielectric slab cavity, and also
calculate the normalization factor $\eta_m(\omega)$ defined in
equation (\ref{eqbiorthdp1}).  In the
cavity interior, we have
\begin{equation}
\varphi_n(x)=\sin \KN
\end{equation}
and
\begin{equation}
\overline{\varphi}^*_m=\sin \KM
\end{equation}
Using the trigonometric identity $\sin(s)
\sin(t)=\frac{1}{2}(\cos(s-t)-\cos(s+t))$, we can write the inner
product (for the case $m \neq n$) as:
\begin{widetext}
\begin{eqnarray}
    && \int_0 ^a \sin\KN \sin\KM dx \nonumber \\
    &&=\frac{\cos\KN \cos\KM}{2n_0}\left(
    \frac{\tan\KN-\tan\KM}{k_n-k_m}-\frac{\tan\KN+\tan\KM}{k_n+k_m} \right) \nonumber \\
    &&=0
    \end{eqnarray}
\end{widetext}
Where we have used the eigenvalue equation (\ref{eqcf_ev}).  The
normalization is:
\begin{equation}
\int_0 ^a \sin^2 \KN dx=\frac{a}{2} \left( 1-\frac{\sin 2\KN}{2\KN}
\right)=\eta_m(\omega)
\end{equation}

\begin{acknowledgments}
We thank Harald Schwefel, T. Harayama and S. Shinohara for useful
conversations; this works was supported by NSF grant
DMR 0408636.
\end{acknowledgments}


\end{document}